\documentstyle[12pt,epsf]{article}

\begin{document}

\begin{flushright}
SLAC--PUB-7205\\
June 1997  
\end{flushright}

\bigskip\bigskip
\begin{center}
{\bf\large
A Short Introduction to BIT-STRING PHYSICS
\footnote{\baselineskip=12pt
Work supported by Department of Energy contract DE--AC03--76SF00515.}
\footnote{\baselineskip=12pt
Conference Proceedings, entitled {\it Merologies}, will be available from
ANPA c/o Prof.C.W.Kilmister, Red Tiles Cottage, Hight Street, Bascombe,
Lewes, BN8 5DH, United Kingdom.}}

\bigskip

H. Pierre Noyes\\
Stanford Linear Accelerator Center\\
Stanford University, Stanford, CA 94309\\
\end{center}
\vfill
\begin{abstract}
This paper starts with a personal memoir of how some significant
ideas arose and events took place during the period from 
1972, when I first encountered Ted Bastin, to 1979,
when I proposed the foundation of ANPA. I then discuss 
program universe, the fine structure paper and its rejection,
the quantitative results up to ANPA 17 and take a new look at the 
handy-dandy formula.
Following this historical material is a first pass at establishing 
new foundations for bit-string physics. 
An abstract model for a laboratory notebook
and an historical record are developed, culminating in the 
bit-string representation. I set up a tic-toc laboratory
with two synchronized clocks and show how this can be used
to analyze arbitrary incoming data. This allows me to
discuss (briefly) finite and discrete Lorentz transformations, 
commutation relations, and scattering theory.
Earlier work on conservation laws in 3- and 4- events and the 
free space Dirac and Maxwell equations is cited. 
The paper concludes with a discussion of the quantum gravity problem
from our point of view and speculations about how a bit-string theory
of strong, electromagnetic, weak and gravitational unification
could take shape.
\end{abstract}
\vfill
\begin{center}
Revised and considerably extended version of \\
two invited lectures presented at the$18^{th}$ annual international meeting of the\\
{\bf ALTRNATIVE NATURAL PHILOSOPHY ASSOCIATION}\\
Wesley House, Cambridge, England, September 4-7, 1996\\
\end{center}
\vfill

\newpage

\section{Pre-ANPA IDEAS: A personal memoir}

\subsection{First Encounters}

When I first met Ted Bastin in 1972 and heard of the Combinatorial Hierarchy
(hereinafter CH),
my immediate reaction was that it must be dangerous nonsense. Nonsense,
because the two numbers computed to reasonable accuracy --- 
$137 \approx \hbar c/e^2$ and $2^{127} + 136 \approx \hbar c/Gm_p^2$ --- 
are {\it empirically determined}, according to
conventional wisdom. Dangerous, because the idea that one can gain
insight into the physical world by ``pure thought'' without empirical
input struck me then (and still strikes me) as subversive of the
fundamental Enlightenment rationality which was so hard won, and which
is proving to be all too fragile in the ``new age'' environment
that the approach to the end of the millennium seems to encourage
\cite{Sagan95,Sokal96}.

Consequently when Ted came back to Stanford the next year
(1973)\cite{Bastin73}, I made sure
to be at his seminar so as to raise the point about empirical input with as much force
as I could. Despite my bias, I was struck from the start of his talk by his obvious
sanity, and by a remark he made early on (but has since forgotten) to the
effect that {\it the basic quantization is the quantization of mass}. 
When his presentation came around to the two ``empirical'' numbers,
I was struck by the thought that some time ago Dyson\cite{Dyson52}
had proved that if one calculates perturbative QED up to the
approximation in which 137 electron-positron pairs can be present,
the perturbation series in powers of $\alpha =e^2/\hbar c \approx
1/137$ is no longer uniformly convergent. Hence, the number 137 {\it as
a counting number} already had a respectable place in the paradigm
for relativistic quantum field theory known as renormalized quantum
electrodynamics (QED). The problem for me became {\it why} should the
arguments leading to CH produce a number which {\it also}
supports this particular physical interpretation.

As to the CH itself, I refer you to Clive Kilmister's introductory talk
in these proceedings\cite{Kilmister97}, where he discusses an
early version of the bit-string construction of the sequence of
discriminately closed subsets with cardinals $2^2-1=3\rightarrow 2^3-1=7\rightarrow
2^7-1=127\rightarrow 2^{127}-1\approx 1.7\times 10^{38}$ based on
bit-strings of length 2,4,16,256 respectively..
The first three terms can be mapped by square matrices of
dimension $2^2=4\rightarrow 4^2=16 \rightarrow 16^2=256$.
The $256^2$ discriminately independent matrices made available by
squaring the dimension needed to map the third level 
are many two few to map the $2^{127}-1$  discriminately closed subsets 
in the fourth level, terminating
the construction. In the historical spirit of this memoir, I add that thanks to some
archeological work John Amson and I did in John's attic in St. Andrews,
the original paper on the hierarchy by Fredrick Parker-Rhodes, drafted
late in 1961, is now available\cite{Parker-Rhodes62}.

I now ask you to join with me here in my continuing investigation of how the CH
can be connected to conventional physics. As you will see in due course,
this research objective differs considerably from the aims of Ted
Bastin and Clive Kilmister. They, in my view, are unnecessarily dismissive
of the results obtained in particle physics and physical cosmology
using the conventional (if mathematically inconsistent) relativistic
quantum field theory, in particular quantum electrodynamics (QED),
quantum chromodynamics (QCD) and weak-electromagnetic unification
(WEU). 

Before we embark on that
journey, I think it useful to understand some of the physics
background. Dyson's argument itself rests on one of the most profound
and important papers in twentieth century physics. In 1937 Carl
Anderson discovered in the cosmic radiation a charged particle he
could show to be intermediate in mass between the proton and electron.
This was the first to be discovered of the host of particles now called collectively
``mesons'' . {\it One} such particle had already been postulated by
Yukawa, a sort of ``heavy photon'' which he showed, using a ``massive
QED'', gave rise to an exponentially bounded force of finite range. 
If the mass of the Yukawa particle was taken to be a few hundred
electron masses, this could be the ``nuclear force quantum''.
Anderson's discovery prompted Gian Carlo Wick to try
to see if the existence of such a particle could be accounted for
simply by invoking the basic principles of quantum mechanics and
special relativity. He succeeded brilliantly, using only one column 
in {\it Nature}\cite{Wick38}. We summarize his argument here.

Consider two massive particles which are
within a distance $R$ of each other during a time $\Delta t$. If they are to act coherently,
we must require $R \leq c\Delta t$. [Note that this postulate, in the
context of my {\it neo-operationalist} approach\cite{Noyes96a}
based on measurement accuracy\cite{Noyes&vdBerginpn}, opens the
door to {\it supraluminal} effects at short distance, which I am now
starting to explore\cite{Noyes96b}]. Because of the uncertainty
principle this short-range coherence tells us that the energy 
is uncertain by an amount $\Delta E\approx \hbar/\Delta t$. 
But then mass-energy equivalence allows a
particle of mass $\mu$ or rest-energy $\mu c^2\geq \Delta E$ to 
be present in the space time-volume of linear dimension $\approx
R\Delta t$. Putting this together, we have the {\it Wick-Yukawa
Principle}:
\begin{equation}
R\leq c\Delta t\approx {c\hbar \over \Delta E} \leq {\hbar/\mu c}
\end{equation}
Put succinctly, if we try to localize two massive particles within 
a distance $R \leq \hbar/\mu c$, then the uncertainty principle allows a
particle of mass $\mu$ to be present. If this meson has the proper
quantum numbers to allow it to transfer momentum between the two
massive particles we brought together in this region, they will
experience a force, and will emerge moving in different directions
than those with which they entered the {\it scattering region}.
Using estimates of the range of nuclear forces obtained from
deviations from Rutherford scattering in the 1930's one can then estimate the mass
of the ``Yukawa particle'' to be $\approx 200-300$ electron masses.

We are now ready to try to follow Dyson's argument. By 1952, one was
used to picturing the result of Wick-Yukawa uncertainty at short
distance as due to ``vacuum fluctuations'' which would allow
$N_e$ electron-positron pairs to be present at distances $r\leq
\hbar/2Nm_ec$. This corresponds to taking $\mu =2Nm_e$ in Eq. 1.   
Although you will not find it in the reference\cite{Dyson52}, in a seminar
Dyson gave on this paper he presented what he called a crude way to understand
his calculation making use of the non-relativistic coulomb potential. 
I construct here my own version of the argument.

Consider the case where there are $N_e$ positive charges in one clump
and $N_e$ negative charges in the other, the two clumps being a
distance $r=\hbar/m_ec$ apart. Then a single charge from one clump will have
an electrostatic energy $N_ee^2/r= N_e[e^2/\hbar c]m_ec^2$ 
due to the other clump and visa
versa. I do recall that Dyson said that the system is dilute enough 
so that non-relativistic electrostatic estimates of this type are
a reasonable approximation. If under the force of this
attraction, these two charges we are considering come together
and scatter producing a Dalitz pair ($e^+  +e^-\rightarrow 2e^+ +2e^-$)
the energy from the fluctuation will add another pair to the system.
Of course this process doesn't happen physically because like charges
repel and the clumps never form in this way. However, in a theory
in which like charges attract [which is equivalent to renormalized QED
with $\alpha_e= [e^2/\hbar c] \rightarrow -\alpha_e$ in the renormalized
perturbation series], once one goes beyond 137 terms such a process  
will result in the system {\it gaining} energy by producing another
pair and the system collapses to negatively infinite energy.
 Dyson concluded that the renormalized perturbation
series cannot be uniformly convergent, and hence that QED cannot be a
fundamental theory, as I have subsequently learned from Schweber's
history of those heroic years\cite{Schweber94}.

Returning to 1973, once I had understood that, thanks to
Dyson's argument,  137 can be interpreted as
a {\it counting number}, I saw immediately that $2^{127}+136\approx
1.7\times 10^{38}\approx \hbar c/Gm_p^2$ could {\it also} be interpreted
as a counting number, namely the number of baryons of protonic mass
which, if found within the Compton wavelength of any one of them, would
form a black hole.
These two observations removed my objection to the
calculation of two pure numbers that, conventionally interpreted,
depend on laboratory measurements using arbitrary units of mass, length
and time. I could hardly restrain my enthusiasm long enough 
to allow Ted to finish his seminar before bursting out with this
insight.  If this cusp turns out to be the point
at which a new fundamental theory takes off --- as I had hoped to make
plausible at ANPA 18 --- then we can tie it firmly into the
history of ``normal science'' as the point where a ``paradigm shift'',
in Kuhn's sense of the word\cite{Kuhn62}, became possible.

However, my problem with {\it why} the calculation made by Fredrick
Parker-Rhodes\cite{Parker-Rhodes62,Parker-Rhodes81a} lead to these
numbers remained unresolved. Indeed, I do not find a satisfactory
answer to that question even in Ted and Clive's book published last
year\cite{Bastin&Kilmister95}. [I had hoped to get further with that quest
at the meeting (ANPA 18, Sept.5-8, 1996), but discussions during 
and subsequent to the meeting  still leave many of my questions
unanswered. I intend to review these discussions and draw my own
conclusions at ANPA 19 (August 14-17,1997).]

\subsection{From ``NON-LOCALITY'' to ``PITCH'': 1974-1979}

My interest in how to resolve this puzzle has obviously continued to this
day. I was already impressed by the quality of Fredrick's results in
1973, and made a point of keeping in contact with Ted Bastin. This led to my
meeting with Fredrick Parker-Rhodes, Clive Kilmister and several of the other Epiphany
Philosophers at a retreat in the windmill at Kings Lynn, followed by discussions in
Cambridge. At that point the group were trying to put together a volume
on {\it Revisionary Philosophy and Science}. I agreed to contribute a
chapter, and finished about half of the first draft of what was to
become ``Non-Locality in Particle Physics'' \cite{Noyes74b,Noyes75} on the
plane going back to Stanford. In the course of finishing that article,
I noted for the first time that the Dyson route to the
hierarchy number places an energy cutoff on the validity of QED at
$E_{max}=2\times 137 m_ec^2$, which is approximately equal to the pion (Yukawa
particle) mass. I have subsequently realized that this explains a
puzzle I had been carrying with me since I was a graduate student.

This puzzle, which I have sometimes called the {\it Joe Weinberg memnonic},
came from quite another direction\cite{JWeinberg47}. An easy way to
remember the hierarchy of nuclear, QED, and atomic dimensions expressed
in terms of fundamental constants is the fact that
\begin{equation} 
1.4 \ Fermi \approx {e^2 \over 2m_ec^2} =\left[{e^2\over \hbar c}\right]{\hbar \over 2m_ec}
=\left[{e^2\over \hbar c}\right]^2{\hbar^2 \over 2m_ee^2} \approx 0.265
 \ Angstrom 
\end{equation}
Why nuclear dimensions should be approximately  half the ``classical
electron radius'' (i.e. ${e^2 \over 2m_ec^2}\approx 1.4\times 10^{-15}
meter$) and hence $[1/137]^2$
smaller than than the radius of the positronium atom (i.e. ${\hbar^2
\over 2m_ee^2}\approx 2.65\times 10^{-10} meter $) was almost completely mysterious
in 1947. It was known that the mass of the electron attributed to it's
electrostatic ``self-energy'' as due to its charge distributed over a spherical shell 
fixed at this radius would have the mass $m_e$, but the success of Einstein's relativity
had shown that this electron model made no sense\cite{Pais82}. The square of this
parameter was also known to be proportional to the cross section for scattering a low energy
electromagnetic wave from this model electron (Thompson cross section
$[8\pi/3](e^2/m_ec^2)^2$),
but again why this should have anything to do with nuclear forces was
completely mysterious. 

As we have already seen, it {\it was} known that the Wick-Yukawa principle  
\cite{Wick38} accounted roughly for the range of nuclear forces if
those forces were
attributed to a strongly interacting particle intermediate in mass between  
proton and electron. However, the only known particle in that mass range
(the muon) had been shown experimentally to interact with
nuclei with an energy $10^{13}$ times smaller  than the Yukawa theory of nuclear forces
demanded\cite{Conversietal47}.
The Yukawa particle (the pion) was indeed discovered later that year,
but there was still no reason to connect it with the ``classical
electron radius''. Joseph Weinberg left his students to ponder this puzzle.

The trail to the solution of this conundrum starts with
a 1952 paper by Dyson\cite{Dyson52}, despite the fact that 
neither he nor I realized it at the time. 
Two decades later, when I first heard a detailed account of the {\it
combinatorial hierarchy}\cite{Bastin73}, and was puzzled by the problem
of how a counting number (i.e. 137) could approximate a combination of
empirical constants (i.e. $\hbar c/e^2$), I realized that this
number is both the number of terms in the perturbation series
and the number of virtual electron-positron pairs where QED ceases
to be self-contained. But, empirically, $m_{\pi}\approx 2\times 137
m_e$. Of course, if neutral this system is highly unstable due
to $2\gamma$ decay, but if we add an electron-antineutrino or a
positron-neutrino pair to the system, and identify the system with
$\pi^-$ or $\pi^+$ respectively, the system {\it is} stable until
we include weak decays in the model. This suggests that the QED theory
of electrons, positrons and $\gamma$-rays breaks down at an energy
of $[2(\hbar c/e^2) +1]m_ec^2$ due to the formation of charged pions,
finally providing me with a tentative explanation for the Joe Weinberg memnonic.
As noted above, I first presented this speculative idea some time ago
\cite{Noyes74b,Noyes75}.

By the time I wrote ``NON-LOCALITY'', I was obviously committed to
engaging in serious research on the combinatorial hierarchy as part of my
professional activity. Ted was able to get a research contract to spend
a month with me at Stanford. I had hoped that this extended period 
of interaction would give me a better
understanding of what was going on; in the event little progress was made on my
side. By 1978 I had met Irving Stein, and was also struggling to
understand how he could get both special relativity and the quantum 
mechanical uncertainty principle from an elementary random walk. His
work, after much subsequent development, is now available in final form
\cite{Stein96}.

Meanwhile Ted had attended the 1976 Tutzing Conference organized by
Carl Friedrick von Weizsacker and presented a paper on the
combinatorial hierarchy by John Amson. I agreed to accompany
Ted to the 1978 meeting and present a joint paper. I arrived in England to
learn of the startlingly successful successful calculation of the
proton-electron mass ratio, which Ted and I had to discuss and
digest in order to present Fredrick's result \cite{Bastinetal79,Parker-Rhodes81b}
at the Tutzing meeting,
which followed almost immediately
thereafter. This formula has been extensively discussed at
ANPA meetings. It was originally arrived at by assuming that the
electron's charge could come apart, as a statistical fluctuation,
in three steps with three degrees of freedom
corresponding to the three dimensions of space and that the
electrostatic energy corresponding to these pieces could be computed
by taking the appropriate statistical average cut off at the proton
Compton radius $\hbar/m_pc$. The only additional physical input
is the CH value for the electronic charge $e^2=\hbar c/137$. Take $0
\leq x \leq 1$ to be the fractional charge in these units and 
$x(1-x)$ the charge factor in Coulomb's law. Take $0\leq y \leq 1$
to be the inverse distance between the charge fractions in that law in
units of the proton Compton radius. Then, averaging between  these
limits with the appropriate weighting factors of $x^2(1-x)^2$ and
$1/y^3$ respectively, Fredrick's straightforward statistical calculation 
gives    
\begin{equation}
{m_p\over m_e} ={137 \pi\over <x(1-x)><{1\over y}>}=
{137 \pi\over ({3\over 14})[1 + {2\over 7} + {4\over 49}]({4\over 5})}
\end{equation}
At that time the result was within a tenth of a standard deviation of
the accepted value. I knew this was much too good because, for example,
the calculation does not include the effect of the weak interactions.
I was therefore greatly relieved when a revision of the fit to the
fundamental constants changed the empirical value by 20 standard
deviations, giving us something to aim at when we know how to
include additional effects.
  
I also learned from the group during those few days before Tutzing 
that up to that point no one had proved the {\it existence} 
of the combinatorial hierarchy in
a mathematical sense! Subsequent to the Tutzing meeting, thanks to the kind
hospitality of K.V.Laurikainen in Finland, I was able to devote considerable time
to an empirical attack on that problem and
get a start on actually {\it constructing} specific representations of
both the $level \ 2 \longrightarrow level \ 3$ and the 
$level \ 3 \longrightarrow level \ 4$ mappings.

It turned out that neither John Amson's nor our contributions to the Tutzing
conferences, despite promises, appeared in the conference proceedings.
Fortunately we had had an inkling at the meeting that this contingency 
might arise. In the event we were able to turn to David Finkelstein
and write a more careful presentation of the developments up to that
point for publication in the {\it International Journal of Theoretical 
Physics}\cite{Bastinetal79}. The first version, called  ``Physical Interpretation 
of the Combinatorial Hierarchy'' (or PICH for short) still lacked a formal existence
proof, but Clive came up with one; further, he and John Amson (whose
unpublished 1976 Tutzing contribution had been extended and completed to serve as an
Appendix) were able to say precisely in what sense the CH is {\it
unique}. The final title was therefore changed to  ``Physical Interpretation 
and mathematical structure of The Combinatorial Hierarchy''
affectionately known as PITCH. The finishing touches on this paper were
completed at the first meeting of ANPA. This brings my informal
history to the point at which Clive ended his historical sketch in
his first lecture. 

\subsection{ANPA 1: The foundation of the organization}

Although I was obviously putting considerable time into trying to
understand the CH, and the Parker-Rhodes formula for $m_p/m_e$ showed
that there might be more to the physics behind it than the basic coupling
constants, I was by no means convinced that the whole effort
might not turn out in the long run to be an unjustifiable
``numerology''. I therefore, privately, took the attitude that my
efforts should go into trying to derive a clear {\it contradiction}
with empirical results which would prove the CH approach to be wrong.
Then I could drop the whole enterprise and get back to my (continuing) conventional
research, where I felt more at home. I was not the only one
with doubts at this time. Clive told us, years later, that he had been
somewhat afraid to examine the foundational arguments too closely
for fear that the whole scheme would dissolve!

In the spring of 1979 I happened to make the acquaintance of an
investment counselor named Dugal Thomas who was advising 
a large fraction of the private
charitable foundations in the US. He offered to help me with
fundraising if I could put together a viable organization for
supporting Ted Bastin's type of research. I threw together a proposal very
quickly. Dugal located a few prospective donors; like all
subsequent efforts to raise substantial funds for ANPA this initial
effort came a cropper. Soon after that effort started I also
learned that I had received a Humboldt U.S.Senior Scientist award,
giving me the prospect of a year in Germany and some extra cash.
Consequently I felt encouraged to approach Clive to see if he would
serve as treasurer for the proposed organization. Clive agreed to approach Fredrick
to see if he would match the small amount of ``seed money'' I
was prepared to invest in ANPA. [The name and original statement of
purpose came from the proposal I had already written. I intended that
the term ``natural philosophy'' in the name of the organization 
would hark back to the thinkers
at the start of the scientific revolution who were trying to
look at nature afresh and shake themselves loose from the
endless debates of the ``nominalist'' and ``realist'' metaphysicists 
of the schools.] With Fredrick's promise in hand, Clive and I approached
Ted Bastin with the invitation to be the Coordinator, and asked John
Amson to join us as a founding member.

The result of all this was what can be properly called ANPA 1,
which met in Clive's Red Tiles Cottage near Lewes in Sussex in the
early fall of 1979.
John Amson was unable to attend, but endorsed our statement of
purpose (modified by Ted to include specific mention of the CH)
and table of organization. Once these details were in hand
we had a proper scientific meeting, including thrashing out
an agreed manuscript for PITCH. I gave a paper on the
quantum mechanical three and four body problem, which I
was working on in Germany. I noted in particular 
that the three channel Faddeev equations go to the seven
channel Faddeev-Yakubovsky equations when one goes from three
to four independent particles, reminiscent of the CH $1 \longrightarrow
2$ level transition. It is taken a long time to see what the
relationship is between these two facts, but now that I am
developing a ``bit-string scattering theory'' with Ed
Jones\cite{Noyes&Jonesinpn}, this old insight is finding an 
appropriate home.   

\begin{center}
{\bf\Large Selected Topics}
\end{center}

All meetings subsequent to ANPA  1 have been held annually in
Cambridge, England.
Proceedings were prepared for ANPA 7\cite{Noyes85}, and some of the papers
incorporated in a SLAC-PUB\cite{Noyes86}. The ANPA 9 proceedings \cite{Noyes87a}
are available from ANPA West. Proceedings ANPA's 10 to 17 
are available from Clive
Kilmister. This is obviously not the place to attempt the impossible
task of summarizing 16 years of work by more than 20 dedicated people
in a way that would do justice to their varied contributions.
I have therefore chosen to pick a few topics where I still find 
continued discussion both interesting and important.

\section{Program Universe}

\subsection{Origin of Program Universe}

About a decade and a half ago, Clive attempted to improve the clarity
of what Ted has called ``the canonical approach'' 
\cite{Bastin96a}
by admitting into the
scheme a second operation besides the {\it Discrimination}
operation, which had been central to the project ever since John Amson
introduced it \cite{Amson65} and related it to subsequent developments
\cite{Amson79}. Clive called this second operation {\it Generation}
because at that stage in his thinking he saw no way to get the
construction off the ground without generating bit-strings as well as
discriminating them. I think he had in mind at that time a random sequence of G
and D operations, but did not quite know how to articulate it.  
Because Mike Manthey and I were unsure how to construct a specific theory from what we
could understand of this new approach, we decided to
make a simple-minded computer model of the process and see how far it
would lead. The first version \cite{Manthey86} turned out to be
unnecessarily complicated, and was replaced 
\cite{Noyes&McGoveran89} by the version described below in section 2.3.
 
One essential respect in which the construction Mike Manthey and I
turned out differs from the
canonical approach is that we explicitly introduced a random element
into the generation of the bit-strings rather than leaving the
background from which they arise vague. Some physicists, in particular
Pauli, have seen in the random element that so far as proved to be
inescapable in the discussion of quantum phenomena an entrance of the
the ``irrational'' into physics. This seems to me to equate
``rationality'' with determinism. I think this is too narrow a view.
Statistical theories of all sorts are used in many fields besides
physics without such approaches having to suffer from being castigated as irrational.
In particular, biology is now founded on the proposition that evolution
is (mainly) explicable as the natural selection of heritable stability in the
presence of a random background. The caveat ``mainly'' is inserted to 
allow for historical contingencies\cite{Gould97}. 
Even in physics, the idea of a random
``least step'' goes back at least to Epicurus,  and of a least step 
to Aristotle. I would characterize Epicurus 
as an exemplary rationalist whose aim was to help mankind escape from the
superstitious terrors generated by ancient religions. This random
element enters program universe via the primitive function ``flipbit''
which Manthey uses to provide either a zero or a one
by unsynchronized access to a closed circuit that flips these two bits
back and forth between two memory locations.
Before discussing how this routine is used, we need to know a bit more about
bit-strings and the operations by which we combine them.

\subsection{Bit-Strings}

Define a bit-string {\bf a}(a;W) with length W and Hamming measure $a$
by its $W$ ordered elements $({\bf a})_w \equiv a_w \in 0,1$ where
$w \in 1,2,...,W$. Define the Dirac inner product, which reduces
two bit-strings to a single positive integer,  by ${\bf a} \cdot {\bf b}
\equiv \Sigma_{w=1}^W a_wb_w$. Hence ${\bf a}\cdot {\bf a}= a$ and
${\bf b}\cdot {\bf b}= b$. Define {\it discrimination} between
bit-strings of the same length, which yields a third string of the same
length, by $({\bf a} \oplus {\bf b})_w = (a_w-b_w)^2$. Clive and I arrived
at this way of representing discrimination during a session in his office
after ANPA 2 or 3. From this representation the {\it basic bit-string theorem}
follows immediately:
\begin{equation}
({\bf a} \oplus {\bf b})\cdot ({\bf a} \oplus {\bf b})=
a+b-2{\bf a} \cdot {\bf b}
\end{equation}
This equation could provide the starting point for an alternative definition 
of ``$\oplus$'' which avoids invoking the explicit structure used above.

We also will need the {\it null string} ${\bf \Phi}(W)$
which is simply a string of $W$ zeros. Note that ${\bf a}\oplus {\bf a}
={\bf \Phi}(W)$, that $({\bf a}\oplus {\bf a})\cdot 
({\bf a}\oplus {\bf a}) = 0$ and that ${\bf a}\cdot {\bf \Phi}=0$. 
The complement of the null string is
the {\it anti-null string} ${\bf W}(W)$ which consists of $W$ ones
and has the property ${\bf W}\cdot {\bf W} = W$.
Of course ${\bf W}\cdot{\bf \Phi}=0$.

Define {\it concatenation}, symbolized by ``$\Vert$'', for two
string ${\bf a}(a;S_a)$ and ${\bf b}(b;S_a)$ with Hamming measures 
$a$ and $b$ and respective lengths $S_a$ and $S_b$ and which
produces a string of length $S_a+S_b$, by
\begin{eqnarray}
({\bf a}\Vert{\bf b})_s &\equiv& a_s \ \ if \ \  s \ \in 1,2,...,S_a\nonumber\\
                   &\equiv& b_{S_a-s} \ \ if \ \ s \ \in S_a+1,S_a+2,...,S_a+S_b
\end{eqnarray}
For strings of equal length this doubles the length of the string
and hence doubles the size of the bit-string space we are using. For
strings of equal length it is sometimes useful to use the
shorthand but somewhat ambiguous ``product notation'' ${\bf a}{\bf b}$
for concatenation. Note that while ``$\cdot$'' and ``$\oplus$'' are, 
separately, both associative and
commutative, in general concatenation is not commutative even for
strings of equal length, although it is always, separately, associative.   
              
\subsection{Program Universe}

To generate a growing universe of bit-strings which at each step
contains $P(S)$ strings of length $S$, we use an algorithm known
as {\it program universe} which was developed in collaboration with
M.J.Manthey \cite{Manthey86,Noyes&McGoveran89}.
Since no one knows how to construct a ``perfect'' random number 
generator, we cannot start from Manthey's ``flipbit'', and must
content ourselves with a pseudo-random number generator that, to some
approximation which we will be wise to reconsider from time to time,
will give us either a ``0'' or a ``1'' with equal probability.
Using any available approximation to ``flipbit'' and
assigning an order parameter $i \in 1,2,...,P(S)$ to each string
in our array, Manthey\cite{Manthey86} has given the coding for
constructing a routine ``PICK'' which picks out some arbitrary string 
${\bf P}_i(S)$ with probability $1/P(S)$. Then program universe
amounts to the following simple algorithm:    

\begin{quotation}
PICK any two strings ${\bf P}_i(S)$,${\bf P}_j(S)$, $i,j \in 1,2,...,P$
and compare ${\bf P}_{ij}={\bf P_i \oplus P_j}$ with ${\bf \Phi}(S)$.

If ${\bf P}_{ij} \neq  {\bf \Phi}$, adjoin  ${\bf P}_{P+1}:={\bf P}_{ij}$
to the universe, set $P:= P+1$ and recurse to PICK. [This process is
referred to as ADJOIN.]

Else, for each $i \in 1,2,...,P $ pick an arbitrary bit ${\bf a}_i
\in 0,1$, replace ${\bf P}_i(S+1):= {\bf P}_i(S)\Vert {\bf a}_i$,
set $S:= S+1$ and recurse to PICK. [This process is referred to as TICK.]
\end{quotation}

It is important to realize that if we take a snapshot of the universe
of bit-strings so constructed at any time, with the ${\bf P}_i$
written as rows of 0's and 1's in a rectangular array containing $S$
columns, 
there is nothing in the
{\it process} that generated them which distinguishes this universe from
any of the $S!$ other universes of 0's and 1's of this height and
width which could be obtained by using any of the $S!$ possible 
permutations of the columns. In this sense any run of program universe
up to this point could just as well have produced any of these other
universes. The point here is that, since the rows are produced by
discrimination, and the order of the bits is the same for each row, the result is
independent of the order of the bits. Similarly, since the column of 
bits which is adjoined to this block representation just before
$S\rightarrow S+1$ is some (hopefully  good!) approximation to a 
Bernoulli sequence, the probability of it having $k$ 1's and $P(S)- k$
0's is simply $P(S)!/k!(P(S)-k)!$ independent of how the rows are
ordered by the order parameter $i$. That is, even though we have
introduced an order parameter for the rows in order to make it easy to
code the program in a transparent way, this parameter in itself is not
intended to play any role in the physical interpretation of the model. 
At this stage in our argument,
this means that program universe can end up with any one of the
$2^{P(S)}S!$ possible block rectangles containing only 0's and 1's 
of height $P(S)$ and width $S$
with some probability which is presumably calculable.
This probability is relevant when we come to discuss cosmology.
Nevertheless, if we look at the {\it internal structure} 
of some fixed portion of {\it any one} of these universes, the way in which
they are constructed will allow us to make some useful and general
statements. Further, these rectangular blocks of ``0'' 's and ``i'' 's
are {\it tables} and hence have {\it shapes} in the  
precise sense defined by Etter's {\it Link Theory}
\cite{Etter96a,Etter96b,Etter96c,Etter97a,Etter97b}.
I hope to have time to discuss Etter's theory at ANPA 19.

I have called another symmetry of the universes so constructed {\it
Amson invariance} in reference to his paper on the BI-OROBOUROS
\cite{Amson85}. He notes that there is nothing in the discrimination operation
which prevents us from using the alternative representation for discrimination
given by 
\begin{equation}
0  \oplus' 0  = 1;\ 0 \oplus' 1=1=1 \oplus' 0;\ 1 \oplus' 1=1
\end{equation}
This will produce a dual representation of the system in which the
roles of the {\it bits} ``0'' and ``1'' (which obviously can no longer
be thought of as integers in a normal notation) are interchanged.
Then when the construction of the {\it combinatorial hierarchy}
is completed at level 4, one will have the complete system
and its dual. But then, one can answer the question which
has been asked in these meetings: ``Where do the bits in the CH come
from?'' in an interesting way. In John's construction the bits are
simply the two dual representations of the CH! Consequently one has
a nested sequence of CH's with no beginning and no end. 
The essential point for me here is not this nested
sequence ---- which will be difficult to put to empirical test ----
but the emphasis it gives to the fact that the two symbols
are {\it arbitrary} and hence that their interchange is a {\it symmetry
operation}. This has helped me considerably in thinking
about how the particle-antiparticle symmetry and CPT invariance
come about in bit-string physics.

Note that in the version of program universe presented here the arbitrary bits are
concatenated only at one growing
end of the strings. Consequently, once the string length
$S$ passes any fixed length $L$
the $P(L)$ strings present will consist of some number
$n_L \leq L$ of strings
which are discriminately independent.  Further, once $S > L$,
the portion of all string of length $L$ changes
only by discrimination between members of this collection.
Consequently it can end up containing at most $2^{n_L}-1$ types of distinct,
non-null strings no matter how much longer program universe runs.
Whether it ever even reaches this bound, and the value of $n_L$ itself,
are {\it historically contingent} on which run of program universe
is considered. This observation provides a model for {\it context sensitivity}.
One result of this feature of program universe is that 
at any later stage in the evolution of the
universe we can always separate any string into two portions,
a {\it label string} ${\bf N}_i(L)$ and a {\it content string}
${\bf C}_i(S-L)$ and write
$P_i(S) ={\bf N}_i(L)\Vert {\bf C}_i(S-L)$ with $i \in 1,2,....,n_L$,
making the context sensitivity explicit..
Once we separate labels from content, the permutation invariance we talked about
above can only be applied to the columns in the label and/or to the
columns in the content parts of the strings separately.
Permutations which cross this divide will interfere with any
physical interpretation of the formalism we have established up to that
point.

In preparation for a more detailed discussion 
on the foundations of bit-string physics, we note here that the alternatives 
ADJOIN and TICK correspond {\it precisely}
to the production of a virtual particle represented by a 3-leg ``Feynman'' diagram,
or ``3-event'', and to the scattering process represented by a 4-leg ``Feynman''
diagram, or ``4-event''respectively. 
We have to use quotes around Feynman, because our
diagrams obey finite and discrete conservation laws consistent
with measurement accuracy. This whole subject will be more
fully developed elsewhere, for instance in discussing
bit-string scattering theory \cite{Noyes&Jonesinpn}.

Another aspect of program universe is worth mentioning. We note that
TICK has a {\it global} character since a 4-event anywhere in the
bit-string universe will necessarily produce a ``simultaneous''
increase of string length in our space of description. This means
that it will be a candidate for representing a coherent cosmological time in
an expanding universe model. The time interval to which TICK
refers is the shortest meaningful (i.e. finite and discrete)
distance that the radius of the universe can 
advance in our theory divided by the velocity of light. We will return
to this idea on another occasion when we discuss cosmology.

\section{Lessons from the rejection of the Fine Structure paper}

\subsection{Background}

In preparation for ANPA 9, Christoffer Gefwert\cite{Gefwertetal87},
David McGoveran\cite{McGoveran&Noyes87} and I\cite{Noyes87b} prepared
three papers intended to present a common philosophical and methodological 
approach to discrete physics. Unfortunately, in order to get the 
first two papers typed and processed by SLAC, I had to put my name on
them, but I want it in the record that my share in Gefwert's and
McGoveran's papers amounted mainly to criticism; I made no substantial
contribution to their work. 
We started the report on ANPA 9\cite{Noyes87a}
with these three papers, followed by a paper on Combinatorial Physics
by Ted Bastin, John Amson's
Parker Rhodes Memorial Lecture (the first in this series), a second
paper by John, and a number of first rate contributed papers. Clive
Kilmister's concluding remarks closed the volume. I went to
considerable trouble to get the whole thing into camera ready format
and tried to get the volume
into the Springer-Verlag lecture note series, but they were unwilling
to accept such a mixed bag. They were interested in the first three
papers and were willing to discuss what else to include, but I was
unwilling to abandon my comrades at ANPA by dropping any of their
contributions to ANPA 9. We ended up publishing  the proceedings 
ourselves, with some much needed help on the physical production 
from Herb Doughty, which we gratefully acknowledge.

David and I did considerably more work on my paper, and I tried to get
it into the mainstream literature, but to no avail. Our joint version
ended up in {\it Physics Essays}\cite{Noyes&McGoveran89}. In the
interim David had seen how to calculate the fine structure 
of hydrogen using the discrete and combinatorial approach, and
presented a preliminary version at ANPA 10\cite{McGoveran88}. I was so
impressed by this result (see below) that I tried to get it
published in {\it Physical Review Letters}. It was rejected even after
we rewrote it in a vain attempt to meet the referee's objections.
In order for the reader to form his own opinion about this rejection,
I review the paper\cite{McGoveran&Noyes91} here and quote extensively from it.  

The first three pages of the paper reviewed the arguments leading to CH
and the essential results already achieved. These will already
be familiar to the careful reader of the material given above. With this as
background we turned to the critical argument:

\begin{quote} 
We consider a system composed of two masses, $m_p$ and $m_e$ |
which we claim to have computed from first principles\cite{Noyes90} 
in terms of $\hbar, c$ and $G_{[Newton]}$ | and identified by their labels using
our quantum number mapping onto the combinatorial hierarchy
\cite{Noyes&McGoveran89}.
In this framework, their mass ratio
(to order $\alpha ^3$ and $(m_e/m_p)^2$)
has also been computed using only $\hbar, c$ and 137.
However, to put us in a situation more analagous to that of Bohr,
we can take $m_p$ and $m_e$ from experiment, and treat $1/137$
as a counting number representing the coulomb interaction;
we recognize that corrections of the order of the square of this number
{\it may} become important one we have to include degrees
of freedom involving electron-positron pairs.
We attribute the binding of $m_e$ to $m_p$ in the hydrogen atom
to coulomb events, i.e. only to those events which involve a specific
one of the 137 labels at level 3 and hence
occur with probability $1/137$;
the changes due to other events average out (are {\it indistinguishable}
in the absence of additional information).
We can have
any periodicity of the form $137 j$ where $j$ is any positive
integer.
So long as this is the only periodicity,
we can write this restriction as $137 j$ {\it steps}
$= 1$ {\it coulomb event}.
Since the internal frequency $1/137j $ is generated independently from
the {\it zitterbewegung} frequency which specifies the mass scale,
the normalization condition
combining the two must be in quadrature. We meet the bound
state requirement that the energy E be less than the system
rest energy $m_{ep} c^2$ (
where $m_{ep}= m_em_p/(m_e +m_p)$ is used to take account of 3-momentum
conservation) by requiring
that $(E/\mu c^2)^2[1 + (1/137N_B)^2]=1$.
If we take $e^2/\hbar c= 1/137$, this is just the relativistic
Bohr formula\cite{Bohr15} with $N_B$ the principle quantum number.
\end{quote}

[Here I inserted into McGoveran's argument a discussion of the 
Bohr formula and how it might
be derived from dispersion theory. This insertion was motivated
by the vain hope that any referee would see that our reasoning was in fact
closely related to standard physics. We will look at this result,
called the handy-dandy formula, in a new way
in the section of this paper carrying that title.]

\begin{quotation}
The Sommerfeld model for the hydrogen atom
(and, for superficially different but profoundly similar reasons
\cite{Biedenharn83}, the Dirac model as well)
requires two {\it independent} periodicities.
If we take our reference period $j$ to be integer
and the second period $s$ to differ from an integer
by some rational fraction $\Delta$, there will be two minimum
values $s_0^{\pm} = 1 \pm \Delta$, and other values of $s$ will
differ from one or the other of these values by integers: $s_n=n+s_0$.
This means that we can relate (``synchronize'') the fundamental
period $j$ to this second period in two different ways, namely to
\begin{equation}
137 j {steps \over (coulomb \ event)}
+137 s_0 {steps \over (coulomb \ event)}
= 1+e=b_+\nonumber
\end{equation}
or to
\begin{equation}
137 j {steps \over (coulomb \ event)}
-137 s_0{steps \over (coulomb \ event)}
= 1-e=b_-\nonumber
\end{equation}
where $e$ is an event probability.
Hence we can form
\begin{equation}
a^2=j^2-s_0^2=(b_+/137)(b_-/137)=(1-e^2)/137^2
\end{equation}
Note that if we want a finite numerical value for
$a$, we cannot simply take a square root,
but must determine from context which of the symmetric factors
[i.e. $(1-e)$ or $(1+e)$]
we should
take (c.f. the discussion about factoring a quadratic above).
With this understood, we write $s_n=n+\sqrt{j^2 -a^2}$.
\par
We must now compute the probability $e$ that $j$ and $s$ are
mapped to the same label, using a
single basis representation constructed within the combinatorial
hierarchy.  We can consider the quantity $a$ as an event
probability corresponding to an event {\bf A} generated by a global
ordering operator which ultimately generates the entire structure
under consideration.  Each of the two events $j$ and $s$ can be
thought of as derived by sampling from the same population.  That
population consists of 127 strings defined at level three of the
hierarchy.  In order that $j$ and $s$ be independent, at least the
last of the 127 strings generated in the construction of $s$ (thus
completing level three for $s$) must not coincide with any string
generated in the construction of $j$.  There are 127 ways in which
this can happen.

There is an additional constraint.  Prior to the completion
of level three for $s$, we have available the $m_2 = 16$ possible
strings constructed as a level two representation
basis to map (i.e. represent)
level three.
One of these is the null string and cannot be used, so there are
15 possibilities from which the actual construction of the label for
$s$ that
completes level 3 are drawn.
The level can be completed
just before or just after some $j$ cycle is completed.
So, employing the usual frequency theory of probability,
the expectation $e$
that $j$ and $s$ as constructed
will be indistinguishable is $e=1/(30\times 127)$.
\par
In accordance with the symmetric factors $(1-e)$ or $(1+e)$
the value $e$ can either subtract from or add to the probability of
a coulomb event.
These two cases
correspond to two different combinatorial paths by which the
independently generated sequences of events may close
(the ``relative phase'' may be either positive or negative).
However we require only the probability that all $s_0$ events
be generated within one period of $j$, which is $1-e$.
Hence the difference between $j^2$ and $s^2$ is to be computed as
the ``square'' of this ``root'', $j^2-s_0^2=(1-e)^2$.
Thus, for a system dynamically bound by the coulomb interaction
with two internal periodicities, as in the
Sommerfeld or Dirac
models for the hydrogen atom,
we conclude that the value of the fine structure constant to be used
should be
\begin{equation}
{1 \over a} = {137\over 1- {1 \over 30 \times 127}} =
137.0359 \ 674...\nonumber
\end{equation}
in comparison to the accepted empirical value of\cite{Aguilar88}
\begin{equation}
{1 \over \alpha} \simeq 137.0359 \ 895(61)\nonumber
\end{equation} 

Now that we have the relationship between $s,j$ and $a$, we consider
a quantity $H'$ interpreted as the energy attribute expressed in
dynamical variables at the $137j$ value of the system containing
two periods.
We represent  $H' $ in units of the invariant system
energy $\mu c^2$.
The independent additional energy due to the shift of
$s_n$ relative to $j$
for a period can then be given as a fraction of this energy by
$(a/s_n)H'$, and can be added or subtracted, giving us the two
factors $(1-(a/s_n)H')$ and $(1+(a/s_n)H')$.
These are to be multiplied
just as we multiplied the factors of $a$ above,
giving the (elliptic)
equation $(H')^2/(\mu¢2c^4) +(a^2/s_n^2)(H')^2/\mu ^2c^4=1$,
Thanks to the previously derived expression of $s=n+s_0$ this
can be rearranged to give us the Sommerfeld formula\cite{Sommerfeld16}
\begin{equation}
H'/\mu c^2=[1 + {a^2 \over (n +\sqrt{j^2-a^2})^2}]^{{-1/2}}\nonumber
\end{equation}

Several corrections to our calculated value for $\alpha$
can be anticipated,....
\end{quotation}

\subsection{The rejection}

It is obvious to any physicist that if an {\it understandable} theory
can be constructed which allows one to calculate the fine structure
constant and Newton's gravitational constant to high accuracy, it
should be possible to create a major paradigm shift in theoretical
physics. But even though David
McGoveran\cite{McGoveran88,McGoveran&Noyes91} 
had showed us how to add four
more significant figures to the calculation of the inverse fine
structure constant, we were unable to make the chain of thought
understandable to most of the physics community. To quote an anonymous
referee for {\it Physical Review Letters}:

\begin{quote}
I recommend that this letter be rejected. How happy we should all be to
publish a physical theory of the fine structure constant! Any such
theory, right or wrong, would be worth publishing. But this letter does
not contain a theory which might be proved right or wrong. The formula
for the fine-structure constant comes out of a verbal discussion which
seems to make up its own rules as it goes along. Somewhere underlying
the discussion is a random process, but the process is never precisely
defined, and its connection to the observed quantities is not
explained. I see no way by which the argument of this letter could be
proved wrong. Hence I conclude that the argument is not science.
\end{quote}   

It should be obvious already that, because of my professional
background, I have some sympathy with this criticism. In fact, though I
was careful not to discuss this with the people in ANPA, for a number
of years my research into the meaning of the CH was, in a sense, aimed at giving
the canonical ANPA arguments sufficient precision so that they {\it could} be
proved wrong. Then, I could drop my involvement with these ideas and
get back to doing more conventional physics. What convinced me that
ANPA must be on the right track was, in fact, the McGoveran calculation
and his later extension of the same ideas to yield several more
mass ratios and coupling constants in better agreement with
experiment\cite{McGoveran89}. Only this past year have we succeeded
in getting two publications about discrete physics into leading
mainstream physics journals\cite{Kauffman&Noyes96a,Kauffman&Noyes96b}. 
But the basic canonical calculations are, even today, not in
the kind of shape to receive that blessing. This is not
the place to review my disheartening attempts to get this and other
ANPA calculations before a broader audience. As an illustration of this
failed strategy of hoping that the quality of our results would do the
job by itself I remind you in the next section what these results were.

\section{Quantitative Results up to ANPA 17}. 

{\bf We emphasize that the only experimental constants needed
as input to obtain these results are $\hbar, c$ and $m_p$.}

The bracketed rational fraction corrections given in bold face type are due
to McGoveran\cite{McGoveran89}. The numerical digits given in bold
face type emphasize remaining discrepancies between calculated and
observed values.

Newton's gravitational constant (gravitation):

$$G_N^{-1} {\hbar c \over m_p^2}=
[2^{127} + 136]\times
{\bf [1-{1\over 3\cdot 7\cdot 10}]} =1.693 \ 31\ldots\times 10^{38}$$
$$ experiment =1.693 \ {\bf 58}(21) \times 10^{38}$$

The fine structure constant (quantum electrodynamics):

$$\alpha ^{-1} (m_e)=
137\times{\bf [1- {1 \over 30 \times 127}]^{-1}} =
137.0359 \ {\bf 674....} $$
$$  experiment =137.0359\ 895(61)$$

The Fermi constant (weak interactions --- $\beta$-decay): 

$$G_Fm_p^2/\hbar c
= [256^2\sqrt{2}]^{-1}\times {\bf [1 - {1 \over 3\cdot 7}]}
=1.02 \ {\bf 758\ldots}\times 10^{-5}$$
$$ experiment = 1.02 \ 682(2)\times 10^{-5}$$

The weak angle (gives weak electromagnetic unification, the $Z_0$ and
$W^{\pm}$ masses).

$$sin^2\theta _{Weak}=
0.25{\bf [1 - {1 \over 3\cdot 7}]^2}
= 0.2267\ldots$$
$$experiment =0.22{\bf 59}(46)]$$

The proton-electron mass ratio (atomic physics):

\begin{equation}
{m_p\over m_e} ={137 \pi\over <x(1-x)><{1\over y}>}=
{137 \pi\over ({3\over 14})[1 + {2\over 7} + {4\over 49}]({4\over 5})}
 = 1836.15 \ {\bf 1497\ldots} 
\end{equation} 
$$ experiment =1836.15\ 2701(37)$$

The standard model of quarks and leptons (quantum chromodynamics):

The pion-electron mass ratios  

$$m_{\pi}^{\pm}/ m_e
=275{\bf [1 - {2\over 2\cdot 3 \cdot 7\cdot 7}]}
= 273.12 \ {\bf 92\ldots}$$
$$ experiment = 273.12 \ 67(4)$$

$$m_{\pi ^0}/m_e
=274 {\bf [1- {3\over 2\cdot 3 \cdot 7 \cdot 2}]}
=264.2 \ {\bf 143\ldots}$$
$$experiment=264.1 \ 373(6)]$$

The muon-electron mass ratio:

$$m_{\mu}/m_e
=3\cdot7\cdot10[1-{3\over 3\cdot7\cdot10}]= 207$$
$$ experiment = 206.768 \ 26(13)$$

The pion-nucleon coupling constant:

$$G^2_{\pi N\bar N}=
[({2 M_N\over m_{\pi}})^2-1]^{{1\over 2}}=[195]^{{1\over 2}}=13.96....$$
$$experiment = 13.3(3), \ or \ greater \ than \ 13.9 $$

I eventually came to the  conclusion that the only way to 
get the attention of the establishment would be to show, in detail,
that these results can be derived from a finite and discrete
version of relativistic quantum mechanics (it turns out, in a finite
and discrete Foch space) which is compatible with most of
the conventional approach. The rest of the paper is devoted to a sketch
of what I think are constructive accomplishments in that direction.
The next section is a draft of the start of a paper illustrating the new strategy.

\section{The Handy-Dandy Formula}

One essential ingredient missing from current elementary particle physics 
is a {\it non-perturbative} connection between masses and coupling constants.
We believe that one reason that contemporary conventional approaches to
relativistic quantum mechanics fail to produce a {\it simple}
connection between these two basic classes of parameters is that they start 
by quantizing classical, manifestly covariant continuous field
theories. Then the uncertainty principle necessarily produces infinite
energy and momentum at each space-time point. While the renormalization
program initiated by Tomonoga, Schwinger, Feynman and Dyson succeeded
in taming these infinities, this was only at the cost of relying on
an expansion in powers of the coupling constant. Dyson\cite{Dyson52} showed in
1952 that this series cannot be uniformly convergent, killing
his hope that renormalized QED might prove to be a fundamental theory
\cite{Schweber94}. Despite the technical and phenomenological successes
of non-Abelian gauge theories, this difficulty remains unresolved
at a fundamental level. What we propose here
as a replacement is an expansion in {\it particle number} rather than
in coupling constant. The first step in this direction already yields
a simple formula with suggestive phenomenological applications, as we
now show.

We consider a two-particle system with energies $e_a,e_b$ and masses
$m_a,m_b$ which interact via the exchange of a composite state
of mass $\mu$. We assume that the exterior scattering state is in a coordinate
system in which the particles have momenta of equal magnitude $p$
but opposite direction.  The conventional S-Matrix approach starts on
energy-shell and on 3-momentum shell with the algebraic
connections\cite{Barnettetal96}
\begin{eqnarray}
e_a^2-m_a^2&=&p^2=e_b^2-m_b^2\nonumber\\
M^2&=&(e_a+e_b)^2-|\vec p_a +\vec p_b|^2\\
|\vec p|(M;m_a,m_b)&=&{[(M^2-(m_a+m_b)^2)
(M^2-(m_a-m_b)^2)]^{{1\over 2}}\over 2M}\nonumber
\end{eqnarray}
but then requires an analytic continuation in $M^2$ off mass shell.
Although this keeps the problem finite in a sense, it leads
to a non-linear self-consistency or {\it bootstrap} problem from which
a systematic development of {\it dynamical} equations has yet to
emerge.

We take our clue instead from non-relativistic multi-particle scattering theory
\cite{Faddeev60,Weinberg63a,Weinberg63b,Weinberg64,Faddeev65,Altetal67,Yakubovsky67} 
in which once a two-particle bound state vertex opens up,
at least one of the constituents must interact with a third particle
in the system before the bound state can re-form. This eliminates the
singular ``self energy diagrams'' of relativistic quantum field theory
from the start. Further, the algebraic structure of the Faddeev
equations automatically guarantees the unitarity of the three particle
amplitudes calculated from them \cite{FLN66}. The proof only requires
the unitarity of the two-body input\cite{Noyes74a,Noyes82}. 
This suggests that it might be possible to develop an ``on-shell''
or ``zero range'' multi-particle scattering theory starting from
some two-particle scattering amplitude formula which guarantees s-wave 
on-shell unitarity.

In order to implement our idea, rather than use Eq.15 we define the
parameter $k^2$, which on shell is the momentum of either particle
in the zero 3-momentum frame, in terms of the variable $s$ which in the physical
region runs from  $(m_a+m_b)^2$ (i.e. elastic scattering threshold)
to the highest energy we consider by
\begin{equation}
k^2(s;m_a+m_b) = s-(m_a+m_b)^2
\end{equation}
Then we can insure on-shell unitarity for the scattering amplitude
$T(s)$ with the normalization $Im \ T(s) =\sqrt{s-(m_a+m_b)^2}|T|^2$ in the
physical region by 
\begin{eqnarray}
T(s)&=&{e^{i\delta (s)}sin \ \delta (s)\over\sqrt{s-(m_a+m_b)^2}}
 ={1\over k\ ctn \ \delta (s) -i\sqrt{s-(m_a+m_b)^2}}\nonumber\\
&=&{1\over \pi}\int_{(m_a+m_b)^2}^{\infty}ds'{\sqrt{s'-(m_a+m_b)^2}|T(s')|^2
\over s' -s-i\epsilon}= 
{2\over \pi}\int_0^{\infty}dk'{sin^2\delta(k')\over k^2 -(k')^2-i\epsilon}
\end{eqnarray}

We arrived at this way of formulating the two-body input for
multi-particle dynamical equations in a rather circuitous way.
It turns out that this representation does indeed lead to well defined
and soluble {\it zero range} three and four particle equations of the 
Faddeev-Yakubovsky
type\cite{Noyes82,Noyes&Jonesinpn}, and that {\it primary singularities} corresponding
to bound states and CDD poles \cite{CDD} can be introduced and fitted
to low energy two particle parameters without destroying the unitarity
of the three and four particle equations. However, if we adopt  the S-Matrix 
point of view 
which suggests that elementary particle exchanges should appear in this
non-relativistic model as ``left hand cuts'' starting at $k^2 = -
\mu_x^2/4$, where $\mu_x$ is the mass of the exchanged quantum
\cite{Noyes&Wong59}, then we discovered\cite{Noyes82} that
the unitarity of the 3-body equations can no longer be
maintained; our attempt to use this model as a starting point for
doing elementary particle physics was frustrated.

We concluded that a more fundamental approach was required, in the
pursuit of which\cite{Bastinetal79,Noyes&McGoveran89} the non-perturbative 
formula which is the subject of this paper was
discovered\cite{McGoveran&Noyes91}. However the reasoning was
considered so bizarre as, according to one referee, not even to
qualify as science. This paper aims to rectify that deficiency
by carrying through the derivation in the context of a relativistic 
scattering theory, which we will call {\it T-Matrix theory} in order
to keep it distinct from the more familiar S-Matrix theory from which it
evolved. Thanks to a comment by Castillejo\cite{Castillejoc90}
in the context of our treatment of the fine structure of the spectrum
of hydrogen\cite{McGoveran&Noyes91},
we finally realized that the success of our new approach required us
from the start to view our {\it T-Matrix} as embedded in a
multi-particle space. This can be accomplished using the relativistic
kinematics of Eq.16 rather than of Eq.15 for the off-shell extension which leads to
dynamical equations.

As we know from earlier work on partial wave dispersion
relations\cite{Noyes&Wong59}, if we know that the scattering
amplitude has a pole at $s_{\mu}=\mu^2$, or equivalently
at $k^2+\gamma^2=0$ where $\gamma =+\sqrt{(m_a+m_b)^2-\mu^2}$
then a subtraction in the partial wave dispersion relation given by
Eq.17 easily accommodates the constraint while preserving on-shell
unitarity in the physical region. This allows us to define the
dimensionless {\it coupling constant} $g^2$ as the ``residue at the
bound state pole'' with appropriate normalization. We choose to
do this by the alternative definition of $T(s)$ given below:
\begin{eqnarray}
T(s;g^2,\mu^2) &=& {g^2\mu\over s-\mu^2}={g^2\mu\over
k^2(s)+\gamma^2}\nonumber\\
&=&{1\over k\ ctn \ \delta(s) +ik(s)} 
\end{eqnarray}
Consistency with the dispersion relation, assuming a constant value for
$g^2$, then requires that at $k^2=0$
\begin{eqnarray}
T((m_a+m_b)^2);g^2,\mu^2) &=& {1\over \gamma}={g^2\mu\over \gamma^2}\nonumber\\
k\ ctn \ \delta((m_a+m_b)^2)&=& \gamma 
\end{eqnarray}
Consequently $g^2\mu=\gamma$ and by taking $\gamma ^2$ also from Eq. 190
we obtain our
desired result, the {\it handy-dandy formula} connecting masses and
coupling constants: 
\begin{equation}
(g^2\mu)^2 =(m_a+m_b)^2 - \mu^2
\end{equation}

In the non-relativistic context where
$\gamma_{NR}^2=2m_{ab}\epsilon_{ab}$, $m_{ab}=m_am_b/(m_a+m_b)$,
$\epsilon_{ab} =m_a+m_b-\mu$, this evaluation of the value of
$k \ ctn \ \delta$ at low energy is equivalent to assuming that
the phase shift is given by the {\it mixed effective range
expansion}\cite{Noyes72}:
\begin{equation}
k \ ctn \ \delta = \gamma + k^2/\gamma= -\gamma +(k^2 +\gamma^2)/\gamma
\end{equation}
corresponding to the {\it zero range} bound state wave function $r\psi (r)=
e^{-\gamma r}$ which assumes its asymptotic form very close to point
where the positions of the two particles coincide. As Weinberg
discusses in considerable detail in his papers on the quasi-particle approach 
\cite{Weinberg63a,Weinberg63b,Weinberg64}, this constraint requires
the bound state to be purely composite --- i.e. to contain precisely
two particles with no admixture of effects due to other degrees of
freedom. We believe that his analysis supports our contention that we can claim
the same interpretation for our relativistic model of a bound state,
and hence that we have derived the proper two-particle input for relativistic
dynamical n-particle equations of the Faddeev-Yakubovsky type. These equations,
which are readily solved for three and four particle systems, will
be presented on another occasion\cite{Noyes&Jonesinpn}.

What follows next is an unsystematic presentation of results,
some of which were initially obtained using the {\it combinatorial
hierarchy}\cite{Bastinetal79,Noyes&McGoveran89}, but which we now 
claim to have placed firmly within
at least the phenomenology of standard elementary particle physics.....

\begin{center}
{\bf\Large THE TIC-TOC LABORATORY: A Paradigm for Bit-String Physics}
\end{center}

Just prior to ANPA 19 I will be attending a conference organized by
Professor Zimmermann entitled {\it NATURA NATURANS: Topoi of
Emergence}. The following notes are intended to serve as raw
material for my presentation there. Some of these ideas came out of
extensive correspondence I have had with Ted and Clive
following ANPA 18, and owe much to their comments. In particular
the section  6 should be compared to Clive's discussion of a
scientific investigation in his paper in these
proceedings\cite{Kilmister97}.

I would also like to remind you before we start of Eddington's
parable that if we set out to measure the length of the fish in the
sea, and we find that they are all greater than one inch long we 
have the option of concluding (a) that all the fish in the sea
are greater than one inch long or (b) that we are using a net with a one inch mesh.
Thinking of my approach in this way, I seem to be finding out that 
{\it because} I insist on finite and discrete measurement accuracy
together with standard methodological principles. I am bound
to end up with something that looks like a finite and discrete
relativistic quantum mechanics that has the ``universal constants''
we observe in the laboratory. Whether the cosmology we observe
is also constrained to the same extent is an interesting question.
My guess is that we will find that historical contingency
plays a significant role. 

\section{A Model for Scientific Investigation}

I restrict our formalism so that it can serve as an abstract
model for physical measurement in the following way. 

We assume that we encounter entities one at a
time, save an entity so encountered, compare it with the next entity
encountered, decide whether they are the same or different, and record
the result. If they are the same, we record a ``0'' and if they
are different we record a ``1''. The first of the two entities
encountered is then discarded and the second saved, ready to be
compared to the next entity encountered. The recursive pursuit
of this investigation will clearly produce an ordered string of ``0''
's and ``1'' 's, which we can treat as a bit-string. We further assume
that this record --- which is our abstract version of a {\it laboratory
notebook} --- can be duplicated, communicated other investigators, treated
as the input tape for a Turing machine, cut into segments which
can be duplicated, combined and compared using our bit-string
operations, the results recorded, and so on. 

Our second assumption
is that if we cut this tape into segments of length $N$ and determine
how many such segments have  
the Hamming measure $a$, the probability we will find
the integer $a$, given $N$ will approach $2^{-N}{N!\over a!(N-a)!}$
in the sense of the {\it law of large numbers}. Without further tests all
such strings characterized by the two integers $a\leq N$ will be
called {\it indistinguishable}. It should be obvious that I make this
postulate in order to be able to, eventually, derive the Planck
black body spectrum from my theory. Remember that Planck's formula has 
stood up to all experimental tests for 97 years, a remarkable
achievement in twentieth century physics! We have recently learned that
in fact it also represents
to remarkable precision the cosmic background radiation at $2.73 ^oK$.
For those who want to know how and why
the quantum revolution started with the discovery of Planck's formula, rather
than just myths about what happened, I strongly recommend Kuhn's last major
work\cite{Kuhn78}.
 
Any further structure coming out of our investigation is to
be found using the familiar operation of {\it discrimination} ${\bf
a}\oplus {\bf b}$ between two strings ${\bf a}$, ${\bf b}$ of equal length, 
by {\it concatenation} ${\bf a}\Vert {\bf b}$ (which doubles
the string length for equal length strings), and  by taking the {\it Dirac inner
product} ${\bf a}\cdot {\bf b}$ which takes two strings out of the category of 
{\it bit-strings} and replaces them
by a positive integer. This third operation is {\it also} how we
determine the Hamming
measure of a single string: ${\bf a}\cdot {\bf a}\equiv a$. It will become
our abstract version of {\it quantum measurement}, which we interpret
as the determination of a {\it cardinal}.

Clearly the category change between ``bit-string'' and ``integer'' is needed
if we are to have a theory of {\it quantitative measurement}.
I take this to be the hallmark of {\it physics} as a science.

The category change produced by taking the inner product
allows us to relate two strings which combine
by discrimination to the integer equation:
\begin{equation}
2{\bf a}\cdot {\bf b}= a+b-({\bf a\oplus b})\cdot ({\bf a\oplus  b})
\end{equation}
If it is taken as axiomatic that (a) we can know the Hamming
measure of a bit string and (b) that this implies that
we can know the Hamming measure of the discriminant between two
bit-strings, then this {\it basic bit-string theorem} 
seems very natural.
  
Once we start combining bit-strings and recording their Hamming
measures, and in particular writing down sequential records of these integers, 
the analysis clearly becomes {\it context sensitive}. It is our
abstract model for a {\it historical record}.

The underlying philosophy
is the assumption that in appropriate units {\it any} physical
measurement can be abstractly represented by a positive integer
with an uncertainty of $\pm {1\over 2}$. If we were using real numbers,
this would be expressed by saying that the value of the physical
quantity represented by $n\pm {1\over 2}$ has a 50\% chance of
lying in the interval between $n-{1\over 2}$ and $n+{1\over 2}$.
But in discrete physics, such a statement is {\it meaningless}
in the sense used by operationalists. Clearly, part of our conceptual
problem is to develop a language describing the uncertainty
in the measurement of integers 
which does {\it not} require us to construct the real numbers.

\section{Remark on Integers}

It is clear from our comment on measurement accuracy that we will 
find it useful to talk about {\it half-integers} as well as integers. This will
also be useful when we come to talk about angular momenta and other
``non-commuting observables'' in our language. But how much farther
must we go beyond the positive integers? It was Kronecker
who said ``God gave us the integers. All else is the work of man.''
One of our objectives is to keep this extra work to a minimum.

I am certain that the largest string length segment we will need to
construct the quantum numbers needed to analyze currently available data about the
observable universe of physical cosmology and particle physics 
is $256$, and that all we need do with such segments is to combine
or compare or reduce them by the operations listed above,
i.e. discrimination, concatenation and inner product.
Using as a basis bit-strings of length $16 W$, I also see how 
to represent negative integers, positive and negative imaginary
integers, and complex integer quaternions. Discussion of how far we
need go in that direction, or into using rational fractions 
other than ${1\over 2}$ is,
in my opinion, best left until we find a crying need to do so.
In any case, we have to lay considerable groundwork before we do.

For the moment we assume all we need know about the integers is that
\begin{equation}
1+0=1=0+1;\ \ \ \ 1\times 0=0=0\times 1;\ \ \ \ 1+1=2
\end{equation}
that we can iterate the third equation to obtain the counting numbers
up to some largest integer $N$ that we pick in advance as adequate
for the purpose at hand, and that given any integer $n$ so generated
other than $0$ or $N$, any second integer $n'$ will be greater than,
equal to, or less than the first. That is, we assume that the three cases
\begin{equation}
n' > n\ \ \ \ or \ \ \ n'=n\ \ \ \ or  \ \ \ n' < n
\end{equation}
are {\it disjoint}. This already implies that we can talk about larger
integers\cite{Kauffman97}.

I have found that McGoveran's phrase ``naming a largest integer 
in advance'', used above, needs be give more structure in my theory.
I assume that all the quantum numbers I need consider 
can be obtained using strings of length 256 or less.
If we have $2^{256}$ such strings, we have more than enough
to {\it count} --- in the sense of Archimedes {\it Sand Reckoner} ---  
the number electrons and nuclei (``visible matter'') in
our universe. The mass of the number of nucleons of protonic mass 
needed to form these nuclei is considerably less
than current estimates of the ``closure mass''
of our universe, leaving plenty of room for the observed ``dark
matter''.  I also believe that considerably less than the  
$256!$ orders in which we could combine the $2^{256}$ distinct strings
of length $256$ will suffice
to provide the raw material for a reasonable model of a historical record of both
cosmic evolution and terrestrial biological, social and cultural
evolution. Such a model can be correct without being complete.  
 
\section{From the Tic-Toc Lab to the Digital Lab}

Our model for an observatory is a number of {\it input} devices which,
relying on our general model, produce bit-strings of arbitrary length
which we can segment, compare, duplicate and operate on using the
three operations ${\bf a}\oplus {\bf b}$, ${\bf a}\Vert {\bf b}$,
and ${\bf a}\cdot {\bf b}$, and record the results. Our model for
a laboratory adds to these observatory facilities, {\it output} devices
which convert bit-strings into signals we can {\it calibrate} in the
laboratory by disconnecting our input devices from the (unknown)
signals coming from outside the laboratory and connecting them to our
locally constructed output devices. We require that the results
correspond to the predictions of the theory which led to their construction.
We then take the critical step from being observers
to being {\it participant observers} by connecting our output devices
to the outside of the laboratory and seeing if the input signals
from the outside into the laboratory
change in a correlated way. Just how we construct input and output
devices is a matter of {\it experimental protocol}, which we must
test by having other observer-participants construct similar devices
and assuring ourselves that they achieve comparable results to our own.
{\it All} of this ``background'' is presupposed in what I mean by
the phrase ``the practice of physics'' which Gefwert, McGoveran and
I employed in our discussion of methodology at ANPA
9\cite{Gefwertetal87,McGoveran&Noyes87,Noyes87b}. It will be seen
that from the point of view of {\it theoretical physics} I am
claiming that all of our operations can, in principle, be reduced to bit-string
operations looking at input tapes to the laboratory,
preparing output tapes connected to the outside world, and 
comparing the new inputs which result from these outputs.
It will sometimes be convenient to refer to these tapes by more
familiar names, such as ``clocks'', ``accumulating counters'', etc., 
without going through the detailed translation
into laboratory protocol that is required for the actual practice
of physics.

The most important device with which to start either an astronomical
observatory or a physical laboratory is a reliable clock.
For us a {\it standard clock} will simply consist of 
an input device which produces a tape with
an alternating sequence of ``1'' 's and ``0'' 's, which
we will also respectively refer to as {\it tic}'s and {\it toc}'s.
It may be a device we construct ourselves or something that occurs
without our intervention other than what is involved in producing the
tape. Note that the fact that we have constructed it still
leaves the physical clock {\it outside} our (theoretical) bit-string
world; it remains essentially just as mysterious as, for example, the pulsations
of a pulsar, as recorded in our observatory. 

We have two types of clock with period $2W$: a {\it tic-toc} clock in which the
sequence of $2W$ alternating symbols starts with a ``1'', and a {\it toc-tic} clock
in which the sequence starts with a ``0''. We will represent the first
by a bit-string which we call ${\bf L}(W;2W)$ and the second by
a bit-string ${\bf R}(W;2W)$. The arbitrariness of the designation
R or L corresponds the fact that our choice of the symbols on
the bit-strings is also arbitrary, reflecting what we called Amson
invariance in our discussion of program universe above. Independent 
of the specific symbolization, these two bit-strings have the following
properties in comparison with each other and with the (unique) anti-null string 
${\bf I}(2W)$ of length $2W$ (which could be called a ``tic-tic clock''): 
\begin{eqnarray}
{\bf R}\cdot {\bf R} =&W& = {\bf L}\cdot {\bf L}\nonumber\\
{\bf R}\cdot {\bf I} =&W& = {\bf L}\cdot {\bf I}\nonumber\\
{\bf R}\cdot {\bf L} =&0&\\ 
{\bf I}\cdot {\bf I} =&2W&\nonumber\\
{\bf R}\oplus {\bf L}=&{\bf I}&\nonumber
\end{eqnarray}

We now have two calibrated clocks one of which we can use to make
measurements, and the second to obtain the {\it redundant} data
which is so useful in checking for experimental error --- a matter
of laboratory protocol which I could expound on at some length,
but will refrain from so doing. We now consider an {\it arbitrary}
signal ${\bf a}(a;2W)$, and compare it with the anti-null string.
We must have either that (1) ${\bf a}\cdot {\bf I}  < W$ or that
(2) ${\bf a}\cdot {\bf I} = W$ or that (3) ${\bf a}\cdot {\bf I} > W$.
Actually, we need consider only the first two cases, because if
we define ${\bf \bar a}$ by the equation ${\bf \bar a}\equiv 
{\bf a}\oplus {\bf I}$, we can reduce the third case to the first
simply by replacing ${\bf a}$ by ${\bf \bar a}$. 
If we know the Hamming measure $a$, for instance by running
the string through an accumulating counter which simply records
the number of ``l'' 's in the string, we do not have to make this test
because, independent of the order of the bits in the string
the definitions of the inner product, the anti-null string ${\bf I}$ 
and the conjugate string ${\bf \bar a}$ guarantee that    
\begin{eqnarray}
{\bf a}\cdot {\bf a} &=& a = {\bf a}\cdot {\bf I}\nonumber\\
{\bf \bar a}\cdot {\bf \bar a} &=& 2W-a = {\bf \bar a}\cdot {\bf I}\\
{\bf a}\cdot {\bf \bar a} &=& 0\nonumber
\end{eqnarray}
Hence an accumulating counter, which throws away most of the
information contained in the bit-string ${\bf a}$, still gives us useful
structural information if we know the context in which
it is employed to produce its single integer result $a$.

If we compare the bit-string ${\bf a}$ with our standard clock {\it before} throwing away the
string, we get two additional integers, only one of which is
independent. Define these by
\begin{equation}
a_R \equiv {\bf a}\cdot {\bf R}; \ \ \ \
a_L \equiv {\bf a}\cdot {\bf L}
\end{equation}
Without actually constructing them, we now know that there exist
in our space of length $2W$ two bit-strings ${\bf a}_R$
${\bf a}_L$ with the following properties
\begin{eqnarray}
{\bf a}_R \oplus {\bf a}_L &=& {\bf a}\nonumber\\
{\bf a}_R \cdot {\bf a}_R &= a_R =& {\bf a}\cdot {\bf R}\nonumber\\
{\bf a}_L \cdot {\bf a}_L &= a_L =& {\bf a}\cdot {\bf L}\\
{\bf a}_R\cdot {\bf L} = 0 = &{\bf a}_R \cdot {\bf a}_L&  
= 0 = {\bf a}_L \cdot {\bf R}\nonumber\\
a_R+a_L &=& a\nonumber
\end{eqnarray}

Suppose we have a second arbitrary string ${\bf b}$ coming from
some independent input device. Clearly we can get some structural
information in the same way as before, succinctly summarized
by the three integers $b, b_R$ and $b_L$ and the 
constraint $b=b_L+b_R$. If the two sources
are {\it uncorrelated}, these amount to a pair of {\it classical
measurements}, which we can, given enough data of the same type,
analyze statistically by the methods developed in classical statistical
mechanics. But if the two sources are correlated {\it and} we 
construct the string ${\bf a} \oplus {\bf b}$ and take its 
inner product both with itself and with our standard clock before
throwing it away, we will have the starting point for a model
of {\it quantum measurement}. This is a deep subject, on which
much light has been shed by Etter's recent papers on Link Theory
\cite{Etter96a,Etter96b,Etter96c,Etter97a,Etter97b}. 
We have started to investigate the connection
to bit-string physics\cite{Noyes97}, but have only scratched the
surface. We trust that the more systematic analysis started in
this paper will, eventually, help in bringing the two together.
Here, we will, instead, show how our tic-toc laboratory can 
give us useful information about the world in which it is embedded.  
   
\section{Finite and Discrete Lorentz Transformations}

We now consider a situation in which our laboratory is receiving
two independent input signals one of which, for segments of length
$2W$, repeatedly gives Hamming measure $a$ and the other $2W - a$. Because of
our experience with the Doppler shift, we leap to the conclusion
that our laboratory is situated between two standard clocks similar
to our own which are sending output signals to us. We assume that
they are at relative rest but that our own lab is moving toward
the one for which the recorded Hamming measure is larger than $W$ and 
away from the second one for which the recorded Hamming measure
is smaller than $W$. We calculate our velocity relative to these
two stationary, signalling tic-toc clocks as $v_{lab} = (a-W)/W$
measured relative to the velocity of light. If our lab is in fact a
rocket ship and we have any fuel left, we can immediately test this
hypothesis by turning on the motors and seeing if, after they have been
on long enough to give us a known velocity increment $\Delta v$,
our velocity measured relative to these external clocks changes to
\begin{equation}
v' = {v +\Delta v\over 1 + v\Delta v}
\end{equation} 
If so, we have established our motion relative to a given, external
framework. Rather than go on to develop the bit-string version
of finite and discrete Lorentz boosts, which is obviously already
implicitly available, I defer that development until we have discussed
the more general bit-string transformations developed in the section
below on commutation relations. For an earlier approach,
see\cite{Noyes92}.

This situation is not so far fetched as might seem at first glance.
Basically, this is how the motion of the earth, and of the solar
system as a whole, have been determined relative to the $2.73 ^oK$ cosmic
background radiation in calibrating the COBE satellite measurements that give us
such interesting information about the early universe.

To extend this ``calibration'' of our laboratory relative to the
universe to three dimensions, we need only find much simpler pairs
of signals than those corresponding to the background radiation,
namely pairs, which for the moment we will call $U$ and $D$ which have
the properties      
\begin{eqnarray}
{\bf U}\cdot {\bf U} &=& W = {\bf D}\cdot {\bf D}\nonumber\\
{\bf U}\cdot {\bf I} &=& W = {\bf D}\cdot {\bf I}\nonumber\\
{\bf U}\cdot {\bf D} &=& 0\\ 
{\bf U}\oplus {\bf D} &=& {\bf I}\nonumber
\end{eqnarray}

These look just like our standard clock, but compared to it we find
that
\begin{eqnarray}
{\bf U}\cdot {\bf R} &=& W +\Delta = {\bf D}\cdot {\bf L}\nonumber\\
{\bf U}\cdot {\bf L} &=& W -\Delta = {\bf D}\cdot {\bf R}
\end{eqnarray}
where (for ${\bf U},{\bf D}$ distinct from ${\bf R},{\bf L}$)
we have that $\Delta \in 1,2,...,W-1$. 
These form the starting point for defining directions and finite and
discrete rotations. As has been proved by McGoveran\cite{McGoveran87},
using a statistical result obtained by Feller\cite{Feller50},
at most three independent sequences which repeatedly have the same number
of tic's in synchrony can be expected to produce such recurrences
often enough to serve as a ``homogeneous and isotropic'' basis for
describing independent dimensions, showing that our tic-toc
lab {\it necessarily} resides in a three-dimensional space.  

It is well known that finite and discrete rotations
of any macroscopic object of sufficient complexity, such as our
laboratory, {\it do not commute}. It is therefore useful
to develop the non-commutative bit-string transformations
before we construct the formalism for finite and discrete Lorentz
boosts {\it and} rotations as a unified theory. 
Once we have done so, we expect to understand better
why finite and discrete commutation
relations imply the finite and discrete version
of the free space Maxwell\cite{Kauffman&Noyes96a} and Dirac
\cite{Kauffman&Noyes96b}equations, which we developed
in order to answer some of the conceptual questions raised by
Dyson's report\cite{Dyson89} and analysis\cite{Dyson90} of Feynman's
1948 derivation\cite{Feynman48} of the Maxwell equations
from Newton's second law and the non-relativistic quantum mechanical
commutation relations; we intend to extend our analysis to gravitation
because we feel that Tanimura's extension of the Feynman derivation
in this direction\cite{Tanimura92} raises more questions than it
answers.

\section{Commutation Relations}

If we consider three bit-strings which discriminate to the null string
\begin{equation}
{\bf a} \oplus {\bf b} \oplus {\bf h}_{ab} = {\bf \Phi} 
\end{equation}
they can always be represented by three {\it orthogonal}
(and therefore {\it discriminately independent})
strings\cite{Noyes95a,Noyes97}
\begin{eqnarray}
({\bf n}_a \oplus {\bf n}_b \oplus {\bf n}_{ab})\cdot   
({\bf n}_a \oplus {\bf n}_b \oplus {\bf n}_{ab}) &=&
n_a+n_b+n_{ab}\nonumber\\
{\bf n}_a \cdot {\bf n}_b &=& 0\nonumber\\
{\bf n}_a \cdot {\bf n}_{ab} &=& 0\\
{\bf n}_b \cdot {\bf n}_{ab} &=& 0\nonumber
\end{eqnarray}
as follows
\begin{eqnarray}
{\bf a} &=& {\bf n}_a \oplus {\bf n}_{ab}\Rightarrow a=n_a+n_{ab}\nonumber\\
{\bf b} &=& {\bf n}_b \oplus {\bf n}_{ab}\Rightarrow b=n_b+n_{ab}\\
{\bf h}_{ab} &=& {\bf n}_a \oplus {\bf n}_b\Rightarrow h_{ab}=n_a+n_b\nonumber
\end{eqnarray}
It is then easy to see that the Hamming measures $a,b,h_{ab}$ satisfy the
triangle 
inequalities, and hence that this configuration of bit-strings can be
interpreted as representing and integer-sided triangle. However, 
if we are given only the three Hamming measures, and invert Eq.35
to obtain the three numbers $n_a,n_b,n_{ab}$, we find that
\begin{eqnarray}
n_{ab} &=& {1\over 2}[+a+b - h_{ab}]\nonumber\\
n_a &=& {1\over 2}[+a-b + h_{ab}]\\
n_b &=& {1\over 2}[-a+b + h_{ab}]\nonumber
\end{eqnarray}
Hence, if either one (or three) of the integers $a,b,h_{ab}$ is (are)
{\it odd}, then $n_a,n_b,n_{ab}$ are {\it half-integers} rather than
integers, and we {\it cannot} represent them by bit-strings.
In order to interpret the angles in the triangle as {\it
rotations}, it is important to start with orthogonal
bit-strings rather than strings with arbitrary Hamming measures.

In the argument above, we relied on the theorem that

{\it if} ${\bf n_i}\cdot {\bf n}_j=n_i\delta _{ij}$ when $i,j \in
1,2,...,N$, 

{\it then}
\begin{equation}
(\Sigma_{\oplus,i=1}^N {\bf n}_i)\cdot (\Sigma_{\oplus,i=1}^N {\bf n}_i)
= \Sigma_{i=1}^N n_i
\end{equation}
which is easily proved \cite{Noyes97}. Thus in the case of two
discriminately independent strings, under the even-odd constraint
derived above, we can always construct a representation of them simply
by concatenating three strings with Hamming measures $n_a,n_b,n_{ab}$.
This is clear from a second easily proved theorem:
\begin{equation}
({\bf a}\Vert {\bf b} \Vert {\bf c} \Vert ....)\cdot
({\bf a}\Vert {\bf b} \Vert {\bf c}....) = a+b+c+....
\end{equation}
Note that because we are relying on concatenation, in order to represent two
discriminately independent strings ${\bf a}$, ${\bf b}$ in this way
we must go to strings of length $W \geq a+b+h_{ab}$ rather than
simply $W \geq a+ b$, as one might have guessed simply from knowing the
Hamming measures and the Dirac inner product. 

If we go to {\it three} discriminately independent strings,
the situation is considerably more complicated. We now need to know
the {\it seven} integers $a,b,c,h_{ab},h_{bc},h_{ca},h_{abc}$,
invert a $7\times 7$ matrix, and put further restrictions on
the initial choice in order to avoid quarter-integers as well as
half-integers if we wish to construct an orthogonal representation
with strings of minimum length $W \geq
a+b+c+h_{ab}+h_{bc}+h_{ca}+h_{abc}$.
We have explored this situation to some extent in the references cited,
but a systematic treatment using the reference system provided by
tic-toc clocks remains to be worked out in detail. 

The problem with non-commutation now arises if we try to get away with
the scalars $a,a_R,\Delta,W$ arrived at in the last section when we
ask for a transformation either of the basis ${\bf R},{\bf L}
\rightarrow {\bf U},{\bf D}$ or the rotation of the string ${\bf a}$
under the constraint $a_R+a_L=a=a_U+a_D$ while keeping the two
sets of basis reference strings fixed. This changes  $a_R-a_L$
to a different number $a_U-a_D$, or visa versa. If one examines this situation
in detail, this is exactly analagous to raising or lowering $j_z$
while keeping $j$ fixed in the ordinary quantum mechanical theory
of angular momentum. Consequently, if one wants to discuss a system in
which both $j$ and $j_z$ are conserved, one has to make a {\it second}
rotation restoring $j_z$ to its initial value. It turns out that,
representing rotations by $\oplus$ and bit-strings then gives
different results depending on whether $j_z$ is first raised and then
lowered or visa versa; finite and discrete commutation relations 
of the standard form result. We will present the details of this
analysis on another occasion. In effect what it accomplishes is
a mapping of conventional quantum mechanics onto bit-strings in such
a way as to get rid of the need for continuum representations
(eg. Lie groups) while retaining {\it finite and discrete} commutation relations.
Then a new look at our recent results on the
Maxwell\cite{Kauffman&Noyes96a} and Dirac\cite{Kauffman&Noyes96b}
equations should become fruitful.

\section{Scattering}

We ask the reader at this point to refer back to our section on the
Handy Dandy Formula when needed. There we saw (Eq. 17)
that the unitary scattering amplitude $T(s)$ for systems  of
angular momentum zero can be computed at
a single energy if we know the phase shift $\delta (s)$ at that energy.
For the following analysis, it is more convenient to work with
the dimensionless amplitude $a(s)\equiv \sqrt{s-(m_a+m_b)^2}T(s)$,which
is related to the tangent of the phase shift by
\begin{equation}
a(s)=e^{i \delta(s)}tan\ \delta (s) = {\tan \delta (s)\over 1 + i\ tan \
 \delta (s)} \equiv {t(s)\over 1 + it(s)}  
\end{equation}
In a conventional treatment, given a real 
interaction potential $V(s,s')=V(s',s)$, $T(s)$ can be obtained by solving
the Lippmann Schwinger equation $T(s,s';z)=V(s,s')$\break
+ $\int ds'' T(s,s'';z)R(s'',s';z)T(s'',s';z)$ with a singular resolvent $R$ and taking
the limit $T(s) = {lim \atop z\rightarrow  s+i0^+,s' \rightarrow s}
T(s,s';z)$. 
Here we replace this integral equation by an {\it algebraic}
equation for $t(s)$: 
\begin{equation}
t(s) = g(s) + g(s)t(s) ={g(s)\over 1 - g(s)}
\end{equation}
One can think of this equation as a sequence of scatterings each
with probability $g(s)$ which is summed by solving the equation.
Here $g(s)$ will be our model of a {\it running coupling constant},
which we assume known as a function of energy. We see that if
$g(s_0)=0$ there is no scattering at the energy corresponding to $s_0$,
while if $g(s_0) = +1$, the phase shift is ${\pi \over 2}$ at the corresponding
energy and $a(s_0) = -i$; otherwise the scattering is finite.

The above remarks apply in the physical region $s> (m_a +m_b)^2$, where
in the singular case a phase shift of ${\pi \over 2}$ causes the
cross section $4\pi sin^2 \delta/k^2$ to reach the unitarity limit
$4\pi \lambda_0^2$ where $\lambda_0=\hbar / p_0$ is the de Broglie
wavelength at that energy; this is called a resonance and the
cross section goes through a maximum value at that energy. If, as in
S-matrix theory, we analyticly continue our equation below
elastic scattering threshold, the scattering amplitude is real
and the singular case corresponds to a bound state pole in
which the two particles are replaced by a single coherent particle
of mass $\mu$, within which the particles keep on scattering
until some third interaction supplies the energy and momentum
needed to liberate them. There can also be a singularity corresponding
to a repulsion rather than attraction, which is called a ``CDD pole''
in S-matrix dispersion theory\cite{CDD}. The corresponding situation in the
physical region is a cross section which 
never reaches the unitarity limit. To cut a long story short,
these four cases correspond to the four roots of the quartic
equation (Eq. 19) called the handy-dandy formula, which we repeat here,
replacing the running coupling constant by its value at the singularity
which we call $g_0=g(s_0)$ 
\begin{equation}
(g_0)^4\mu^2= (m_a+m_b)^2 -\mu^2
\end{equation}
Again to cut a long story short, the model for a running coupling
constant which Ed Jones and I are exploring\cite{Noyes&Jonesinpn}
is simply
\begin{equation}
g_{m_a,m_b;\mu}(s)=\sqrt{ { \pm [(m_a+m_b)^2 -\mu^2](m_a+m_b)^2
\over[k^2(s)-(m_a+m_b)^2]s} }g_{m_am_b;\mu}(0)
\end{equation}
The singularity at $s=0$ is included only when $m_a$ and $m_b$
have a bound state of zero mass, usually called a quantum. 

We have seen that when $m_a=m_e$, $m_b=m_p$ and $\mu = m_H$
the handy-dandy formula gives the relativistic
Bohr formula for the hydrogen spectrum. Replacing $m_p$ by $m_e$
in the formula gives the corresponding formula for positronium (i.e
the bound state of an electron-positron pair). But for that system,
one can think of the photons produced in electron-positron
annihilation as bound states of the pair with zero rest mass.
This interaction is important in high energy electron-positron
scattering, where it is called ``Bhabha scattering''. Introducing
the $s^{-{1\over 2}}$ in this way is supposed to insure that our
theory gives the correct Feynman diagram (and hence cross section)
for this effect, but until we have checked the detailed derivation and
predictions I warn the reader to treat this formula (Eq.42)
as a guess rather than as a result actually derived from the theory. 
 
In  my paper at ANPA WEST 13\cite{Noyes97}, I started to explore the
connections of this type of scattering theory 
to bit-strings {\it and} to Etter's Link Theory by making
the hypothesis that \begin{equation}
tan \delta_{ab} = {\pm ({\bf a}\oplus {\bf b})\cdot 
({\bf a}\oplus {\bf b})\over {\bf a}\cdot {\bf b}}
\end{equation}   
Unfortunately the details are about a sketchy as presented here, but at
least should provide insight into where I am headed.

\section{Quantum Gravity}

The initial intent of Eddington, and following him of Bastin and
Kilmister, was to achieve the reconciliation of quantum mechanics with
general relativity. I emphasize here that they were aware of the
problem, and thought they had a research program which might solve it,
long before the buzz-words ``Grand Unified Theory'', ``Theory of
Everything'', ``Final Theory'' or ``Ultimate Dynamical Theory''
\cite{tHooft94,Noyes95b} became popular. In a sense they achieved the
first major step with the publication of Ted Bastin's paper in
1966\cite{Bastin66}. According to John Amson, that paper came about
after many attempts to understand Fredrick's
breakthrough\cite{Parker-Rhodes62} had led John to the discovery of
discriminate closure\cite{Amson65} which gave more mathematical coherence to the
scheme, and also convinced Clive and Fredrick that it was time to
publish. In the event, the four authors could not agree on a text in
time to meet a deadline, and authorized Bastin to go ahead with his
version as sole author. This paper really does unify electromagnetism 
with gravitation (which was also Einstein's long sought and unachieved
goal) in the sense that both coupling constants are derived
from a common theory. Ted also correctly identified $(256)^4$ with the
weak interactions, but missed the $\sqrt{2}$ needed to connect
it numerically with the Fermi constant because he was unfamiliar with
the difference between 3-vertices (Yukawa-type couplings) and
4-vertices (Fermi-type couplings) in the quantum theory of fields.
As we all know, this paper was met by resounding silence.
I am optimistic, in spite of my past failures, that
the bit-string theory now has enough points of contact with 
more conventional approaches to fundamental physics to get us all
into court.

Since the critical problem for many physicists is how we deal with
``quantum gravity'', I start there. For weak gravitational fields
it makes sense to start in a flat
space-time\cite{Weinberg65}. Then it can be shown
that spin 2 gravitons of zero mass lead to the Einstein field
equations, but as Meisner, Thorne and Wheeler note\cite{Misneretal70},
the ``Resulting theory eradicates original flat geometry from all
equations, showing it to be unobservable.'' Consequently they feel
that this approach says nothing about  ``... the greatest single crisis
of physics to emerge from these equations: complete gravitational
collapse.'' My qualitative answer is that this crisis arises from using a continuum
theory at short distance where only a quantum theory makes sense.
I now try to make the alternative presented here plausible.

My first step is to establish the existence of quantum gravitational effects
for {\it neutral} particles, namely neutrons. That neutrons are
gravitating objects in the classical sense was proved at Brookhaven
soon after the physicists there learned how to extract epithermal neutrons from their
high flux reactor and send them down an evacuated pipe a quarter of a
mile long. The neutrons fell (within experimental error) by just the amount
that Galileo would have predicted. That they are quantum mechanical
objects was proved by Overhauser\cite{Overhauseretal75} by cutting a single silicon crystal 10
centimeters long into three connected planes and using critical
reflection, both {\it calculable} from measured $n-Si$ cross sections
and demonstrable {\it experimentally}, to form in effect
a two-slit apparatus for neutrons with the positions known to atomic
precision over a distance of ten centimeters. Then the shift in the interference
pattern between the case when two beams were both horizontal to
the case when they were in the vertical plane with one higher than the
other for part of its path was proved to be precisely that predicted
by non-relativistic quantum mechanics using the Newtonian
gravitational potential in the Schroedinger equation. It was this
brilliant experiment which convinced me that quantum mechanics is a
general theory and not just a peculiarity in the behavior of
electrically charged particles at short distance. In my opinion,
Overhauser deserves the Nobel prize for this work, which opened
up the study of the foundations of quantum mechanics to high
precision experimental investigation. In the hands of Rausch\cite{Rausch86}
and others this technique has led to many tests of the model of
the neutron as a quantum mechanical particle acting coherently
with a precisely known mass and magnetic dipole. 

Having established that neutral particles react gravitationally to
the Newtonian gravitational potential $V_N(m_1,m_2;r) =G_N{m_1m_2\over
r}$ as expected, it makes sense to extend our relativistic bit-string
model for the Coulomb potential  $V_C(m_1,m_2;r) ={Z_1Z_2e^2\over r}$ 
to the gravitational case.
Here $Z_1,Z_2$ are the electric charges expressed in units of the electronic
charge $e$.  If we are guided by Bastin's remark
quoted above to the effect that the basic quantization is the
quantization of mass, the analogy suggests that there is a (currently
unknown) unit of mass, which we will call $\Delta m$. Then, to complete
the analogy with the Coulomb case, we can replace $\alpha _C=e^2/\hbar
c\approx 1/137$ with a much smaller constant $\alpha_N=G_N{\Delta m^2/\hbar
c}$. If we also define $N_i=m_i/\Delta m$ for any particle $i$ with 
{\it gravitational mass} $m_i$, the quantized version of the Coulomb and
Newtonian interactions become formally equivalent, differing only by two
dimensionless constants and two {\it quantum numbers}, independent of the 
units of charge or mass:
\begin{equation}
V_C(Z_1,Z_2;r) = Z_1Z_2\alpha_C{\hbar c\over r}; \ \ \ \  
V_N(N_1,N_2;r) = N_1N_2\alpha_N{\hbar c\over r}
\end{equation}

We can now apply the Dyson argument to gravitation with more precision.
We have seen that renormalized QED extended to enough precision to generate
$N_e=137$ electron-positron pairs, becomes unstable because of
(statistically rare) clumping  of clusters with enough electrostatic
energy to form another pair. We interpret the fact that this disaster
does not occur to the formation of a pion with mass $m_{\pi} \approx 2\times
137 m_e$. In the neutral particle case, if one assumes CPT invariance,
one can still distinguish fermions from anti-fermions by their
spin even if they have no other quantum numbers. Hence, independent
of whether or how neutral fermions and anti-fermions interact, we can
expect gravitational clumping to occur for each type separately.
Recall Dyson's remark that the system is dilute enough so that
the non-relativistic potential can be used reliably to estimate
the interaction energy of the clump. We know experimentally
that $\Delta m$ is much smaller than the electron mass so that
the critical radius of the clump is $\hbar/\Delta m c >> \hbar/m_ec$,
so Dyson's comment still applies..

In contrast to the electromagnetic case where the cutoff mass-energy
of the pion requires us to go outside QED for the physics,
in the gravitational case we have a cutoff energy ready to hand,
namely the Planck mass $M_{Pk} \equiv [{\hbar c/G_N}]^{{1\over 2}}$.
If we assemble a Planck's mass worth of neutral particles of mass
$\Delta m$ at rest within their own Compton wavelength, i.e. 
$N_G\Delta m=M_{Pk}$, and nothing
else intervenes, they will fall together until they are all within
a distance of $\hbar/M_{Pk}c$. At that point they will have a
gravitostatic energy $N_G G_N\Delta m M_{Pk}/[\hbar/c\Delta
M]=M_{Pk}c^2$, which is just sufficient
to contain the kinetic energy they acquired in reaching this
concentration. Inserting the definition of the Planck Mass
into this gravitational energy equation we find that it is
algebraically equivalent to the boundary condition  with which
we started: $N_G\Delta m=M_{Pk}$.
They will form a black hole with the Planck radius.
We conclude that $N_G=M_{Pk}/\Delta m$ neutral, gravitating
objects of mass $\Delta m$ at rest within their own Compton wavelength will
collapse to a black hole with the Planck radius, a quantum version of
the disaster that Wheeler is concerned about. 

If there are, in fact,
neutral fermions with no other properties than their mass, they would form such
such  black holes and might serve as a model for the dark matter
which we know to be at least ten times as prevalent in the universe as
ordinary matter. It then becomes a question in big-bang cosmology
whether or not they contribute the needed effects to correlate
additional observations. We defer that question to another occasion. 

If we assemble enough particles of the
types we know about to add up to a Planck mass, they  
can start collapsing and will radiate much of their energy on the
way down to higher concentrations. If attractions are
balanced by repulsions they could end up close enough together to form
a black hole. However, as Hawking showed, small black holes
interact with the ``vacuum'' outside the 
event horizon and radiate electromagnetically with the consequence
that they are  ``white hot''
and soon evaporate; the calculation was extended to rotating, charged
black holes by Zurek and Thorne\cite{Zurek&Thorne85}.  
At the quantum scale a new possibility enters, namely that
a quantum number may be possessed by the system which {\it cannot}
be radiated away by emitting a single particle with that quantum number
while conserving energy, momentum and spin. The obvious candidates
for such conserved quantum numbers are baryon number, charge and lepton
number, suggesting that the lightest baryon (the proton), the
lightest charged lepton (the electron) and the lightest neutral lepton
(the electron-type neutrino) are {\it gravitationally stabilized}
black holes with spin ${1\over 2}\hbar$. This idea did not make it
into the mainstream literature\cite{Noyes91}. We use it freely
in what follows. The problem then is to explain why $(M_{Pk}/m_p)^2
\approx 2^{127}$, why $m_p/m_e \approx 1836$ and why $m_{\nu_e}/m_e
\leq 5\times 10^{-5} m_e$.

Before we leave gravitation, however, we need to show within bit-string
physics that the graviton has spin 2. We know from our discussion of
the handy-dandy formula that we can account for spin ${1\over 2}$
electrons,  positrons and protons and their interactions with the
appropriate spin 1 photons. We have shown elsewhere\cite{Noyes94}  
that we can construct the quantum numbers of the standard model of
quarks and leptons. In particular, this will include the electron,
muon and tau neutrinos and their anti-particles. Extending this
approach to a string 
of length 10 we can have 6 spin ${1\over 2}$ fermions and 2 spin 1
bosons. On another occasion I will show how these construct 5 gravitons
and 5 anti-gravitons represented in terms of strings of length 10,
and go on from that to make a model for dark matter that can be expected
to be approximately 12.7 times as prevalent in the universe as electrons
and nucleons. 

It remains to show that we can meet the three classical
tests of general relativity, a problem met on another occasion
\cite{Noyes&McGoveran88}. Briefly, any relativistic theory gives the 
solar red shift (Test 1), the factor of 2 compared to special relativity
in the bending of light by the sun comes from the spin 1 of the photon
(Test 2), and the factor 6 compared to special relativity for the precession
of the perihelion of Mercury\cite{Bergmann42} from the spin 2
of the graviton (Test 3). The calculation by Sommerfeld on which the third
argument partly depends comes from simply replacing the factors
$N_i=m_i/\Delta m$ in the Coulomb potential by $E_i/\Delta m$,
where $E_i$ includes the changing velocity of the orbiting particle 
in elliptical orbits, and hence is natural in our theory.  
  
\section{STRONG, ELECTROMAGNETIC, WEAK, GRAVITATIONAL UNIFICATION
(SEWGU): A look ahead}
  
I now show, very briefly, how SEWGU {\it might} take shape, if all goes
well, giving the zero$^{th}$ approximation to some
old results such as $[\alpha_e^{-1}(0)]_0=137$, some new results I
believe such as $[m_{\tau}/m_{\mu}]_0=16$, and some guesses
such as $[m_t/m_Z]_0=2$ which I have little confidence in.
The notation makes no distinction between fact and speculation,
so {\it CAVEAT LECTOR}!

Start with strings of length 8 to label the 6 L,R {\it bare} states of the
three types of neutrinos $\nu_e,\nu_{\mu},\nu_{\tau}$ and their
anti-particles, together with two slots for three generations,
$g_1=(10),g_2=(11),g_3=(01)$. To get masses and energies we have to add
content strings and do a detailed analysis using bit-string scattering
theory. We expect the following results to emerge
\begin{equation}  
[{m_{\mu}\over m_e}]_0 =210=  [{m_{\nu_{\mu}}\over m_{\nu_e}}]_0
\end{equation}
\begin{equation}
[{m_{\tau}\over m_{\mu}}]_0  =16= [{m_{\nu_{\tau}}\over m_{\nu_{\mu}}}]_0 
\end{equation}
together with the massless bosons $\gamma_L,\gamma_R,\gamma_C$ (two
states of spin 1 photons with the coulomb interaction as a third state)
and $g_{2L},g_{1L},g_0,g_{1R},g_{2R},g_N$ (five states of the graviton
plus the Newtonian interaction). 

The electromagnetic coupling at the mass of the $Z_0$,
$[\alpha_e^{-1}(m_Z)]_0=128$.

The masses of the charged and neutral pion compared to the electron:
\begin{equation} 
[{m_{\pi^{\pm}}\over m_e}]_0  =275 = 2[\alpha^{-1}_e(0)]_0 +1
\end{equation}
\begin{equation}
[{m_{\pi^0}\over m_e}]_0  =274 = 2[\alpha^{-1}_e(0)]_0
\end{equation}

We can also use label strings of length eight to get (bare) quarks with
eight colors\break 
(red,orange,yellow,green,blue,purple,black,white) which
can form the colorless pion triplet ($\pi^+,\pi^0,\pi^-$) and 
the nucleon-antinucleon doublet ($n,p,\bar n, \bar p$).
To identify these as first generation hadrons, we
neead to extend the string length from 8 to 10.
This allows us to go on to strings of length 16 and include the neutrinos
with the two generation slots accounting for the shared coupling.
After a few years of effort, we expect parameters of the full 
Cabbibo-Kobayashi-Maskawa
coupling scheme to emerge. If this fails,
we may have to abandon bit-string physics!

Calling the strong coupling constant $\alpha_{\pi}$ rather than
$\alpha_s$ to emphasize the conceptual difference, we are confident 
that at low energy  
\begin{equation}  
[\alpha_{\pi}(m_{\pi}^2)]_0 = 1
\end{equation}
\begin{equation}  
[\alpha_{\pi}^{-1}(4m_p^2)]_0 = 7
\end{equation}

At high energy we expect to show that
\begin{equation}  
[{m_Z\over m_p}]_0 = 2\times 7^2
\end{equation}
\begin{equation}
[{m_t\over m_Z}]_0 = 2
\end{equation}
where $m_t$ is the top quark mass. If this works out we may be able to
predict a new coupling between quarks and leptons that goes
beyond the standard model which {\it might} explain the anomalous
results recently obtained at ZEUS and HERA.

I expect to be able to derive the $m_p/m_e$ formula in a way consistent
with McGoveran's last paper on that problem\cite{McGoveran90}, but now 
directly using bit-string dynamics.

I expect to be able to understand the mapping between Foch space labels
and bit-string geometry in terms of the Eulerean rectangular block
Kilmister told me about with edges of length 44, 117, 240 using
strings of length 256 divided into a label of length 16 and content
string of length 240.

Finally, I expect to recover the old result
\begin{equation}  
[{M_{Pk}^2\over m_p^2}]_0 = 2^{127}
\end{equation}
in terms of a running coupling constant for gravitation normalized by
$\alpha_G(M_{Pk}^2)=1$ using bit-string scattering theory.
\section{Epilogue}

I realize all too well how sketchy these notes are and apologize for that.
I hope to get a systematic and reasonably complete outline of the full
theory hammered out in a year or two. ``But always at my back I hear
time's winged chariot hurrying near. And yonder all before us lie
deserts of vast eternity.... The grave's a fine and private place, but none
I think do there embrace...'' theoretical physics! So I decided to
rough out this paper to insure that it is part of the Fixed Past before
some singular event in the Uncertain Future terminates my activities.

\end{document}